\documentstyle[epsf,11pt]{article}
\newcommand{\be}{\begin{equation}}
\newcommand{\ee}{\end{equation}}

\newcommand{\cmod}[1] {|{#1}|^2}

\newcommand\ph{\varphi_{\ae}}

\newcommand\pc{\varphi}
\newcommand\pr{\varphi_r}
\newcommand\pl{\varphi_l}
\newcommand\pd{\varphi_d}

\begin{document}
\title{Adiabatic self-trapped states in carbon nanotubes\\
}

\author{
L. Brizhik\thanks{e-mail address: brizhik@bitp.kiev.ua},\,
A. Eremko\thanks{e-mail address: eremko@bitp.kiev.ua},\,
\\
Bogolyubov Institute for Theoretical Physics, 03143 Kyiv, Ukraine\\
B. Piette\thanks{e-mail address: B.M.A.G.Piette@durham.ac.uk},\,
M. Watson\thanks{e-mail address: m.j.watson@durham.ac.uk},\,
and
W. Zakrzewski\thanks{e-mail address: W.J.Zakrzewski@durham.ac.uk}
\\
Department of Mathematical Sciences,University of Durham, \\
Durham DH1 3LE, UK\\
}
\date{}
\maketitle

\begin{abstract}

We study here polaron (soliton) states of electrons or holes in a 
model describing carbon-type nanotubes. In the Hamiltonian of the 
system we take into account the electron-phonon interaction that 
arises from the deformation dependencies of both the on-site and the 
hopping interaction energies.  Using an adiabatic approximation, we 
derive the equations for self-trapped electron states in zigzag 
nanotubes. We find the ground states of an electron in such a system 
and show that the polaron states can have different symmetries 
depending on the strength of the electron-phonon coupling.  Namely, 
at relatively weak coupling the polarons possess 
quasi-one-dimensional (quasi-1D) properties and have an azimuthal 
symmetry.  When the coupling constant exceeds some critical value, 
the azimuthal symmetry breaks down and the polaron spreads out in 
more than one dimension.

We also study polarons that are formed by the electrons in the 
conducting band (or by holes in the valence band) in semiconducting 
carbon nanotubes.  We show that their properties are more complex 
than those of quasi-1D ground state polarons. In particular, polarons 
in semiconducting carbon nanotubes possess an inner structure: being 
self-trapped along the nanotube axis they exhibit some modulations 
around the nanotube.

\end{abstract}

\section{Introduction}

Over the last few years much work has been done on studying physical 
properties of carbon nanotubes \cite{SaiDrDr,DrDrEkl,Dai}, and boron 
nitride nanotubes \cite{Benny}.  The experimental studies of such 
nanosystems have revealed their peculiar properties that are 
important for practical applications \cite{nanosw}.  

Not surprisingly, carbon and boron nitride nanotubes are quite 
complex systems.  Their geometry is based on a deformable hexagonal 
lattice of atoms which is wrapped into a cylinder.  Experimental and 
theoretical studies show the important role of the nanotube geometry: 
many properties of nanotubes can be modified in a controllable way by 
either varying the nanotube diameter and chirality, {\it i.e.} the 
way the lattice is wrapped into a cylinder, \cite{SaiDrDr,DrDrEkl}, 
or  by doping them with impurity atoms, molecules and/or compounds 
\cite{Duc}. Theoretical studies of single wall carbon nanotubes 
(SWNT) \cite{WoMah,Mah} have demonstrated the importance of the 
interaction of electrons with lattice vibrations 
\cite{SFDrDr,MDWh,Kane,Chamon,Alves,WoMah,JiDrDr,PStZ}.  Note  that 
sufficiently  long  SWNTs can be considered as one-dimensional (1D) 
metals or semiconductors depending on their diameter and chirality 
\cite{SaiDrDr,DrDrEkl}.  The nanotubes possess a series of electron 
bands, which can be determined  by 1D energy dispersion relations for 
the wave vector $k$ along the axis of the nanotube.  

In 1D systems the electron-phonon coupling can lead to the formation 
of self-trapped soliton-like states (large polarons) which can move 
with a constant momentum \cite{Dav}. In 1D metals, due to the Peierls 
instability \cite{Peierls}, the energy gap appears at the Fermi level 
and the Fr\"ohlich charge-density wave is formed \cite{Froehlich} 
instead of a soliton.  Recent experiments \cite{Furer,Rados} have 
shown that even long channel semiconductor SWNTs may have very high 
mobilities at a high doping level.  The posibility for the formation 
of states which spontaneously break symmetry in carbon nanotubes  has 
been discussed in \cite{MDWh,Kane,Chamon}.  In particular, large 
polarons (solitons) in nanotubes have recently been studied in 
\cite{Alves,Pryl} where the long-wave approximation has been used for 
the states close to the Fermi level. However, such a description, 
equivalent to the continuum approximation, does not take into account 
some important aspects of the crystallographic structure of the 
system. 

In this paper, first, we consider the ground states of a 
quasiparticle (electron, hole or exciton) in the zigzag nanotube 
system and, second, we study the polaron states of an electron in the 
lowest unfilled (conducting) band or an extra hole (an electron 
deficiency) in the highest filled (valence) band in carbon nanotubes.  
For this we use the semi-empirical tight-binding model with 
nearest-neighbour hopping approximation \cite{SaiDrDr}. The 
advantages of this method for some 1D systems, like polyacetylene and 
 carbon nanotubes,  have been demonstrated in \cite{SSH} and 
\cite{SaiDrDr,SFDrDr,MDWh}, respectively.  We study a quantum system 
involving a hexagonal lattice of atoms and electrons and then perform 
an adiabatic approximation. Then we derive the system of discrete 
nonlinear equations, which as such, can possess localised 
soliton-like solutions. We perform an analytical 
study of these equations and show that, indeed, this is the case, and 
various polaron states can be formed in the system.  In fact, these 
equations were used in \cite{us} to determine numerically the 
conditions for the formation of such polaron states.  Our analytical 
results on self-trapped states of a quasiparticle are in good 
agreement with the results obtained in \cite{us}. 
We also study polarons that are formed by the electrons in the 
conducting band (or by holes in the valence band) in semiconducting 
carbon nanotubes.  

The paper is organised as follows. The next section presents the 
model of the nanotube. The phonon Hamiltonian is discussed in Sect. 
3, electron in Sect. 4 and in Sect. 5 we discuss the electron-phonon 
interactions.  The details of the diagonalization of the electron 
Hamiltonian are presented in Appendix 1.  In Section 6 we determine 
the adiabatic and non-adiabatic terms of the Hamiltonian.  The 
corresponding zero-order adiabatic approximation then leads to the 
equations for the self-trapped electron states while the 
non-adiabatic term of the Hamiltonian provides a self-consistent test 
to determine the conditions of applicability of the adiabatic 
approximation.  The system of equations in the zero-order adiabatic 
approximation in the site representation is derived in Appendix 2.  
In Sect. 7 we derive some analytical solutions for the large polaron 
ground state, and in Sect. 8 we discuss the transition to the states 
with broken axial symmetry. In Section 9 we study large polaron 
states in semiconducting carbon nanotubes. The paper ends with 
conclusions.

\section{Model of a Nanotube}

In this section we define the variables to describe a nanotube. Let $d$ be 
the length of the side of the hexagons of the nanotube, 
$R$ its radius and let $N$ be the number of hexagonal cells wrapped around 
the nanotube. Then we have 
\begin{equation}
\alpha = 2\pi / N,\qquad
a = d \sqrt{3}, \qquad
b = d/2,        \qquad
a = 4R\sin({\alpha\over4}),
\end{equation}
where $a$ is the distance between two next to nearest neighbour sites.

\begin{figure}[htbp]
\unitlength1cm \hfil
\begin{picture}(8,8)
 \epsfxsize=8cm \epsffile{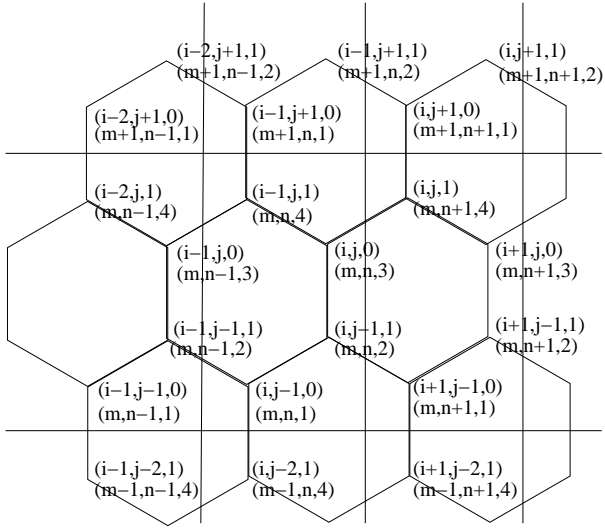}
\end{picture}
\caption{The two index labelling on the Hexagonal lattice.}
\end{figure}

To label all sites on the nanotube one can use
two different schemes. They involve having
2 or 4 lattice sites as a basic unit.
The first one, used in \cite{us}, 
is closely connected with a  unit cell of a graphene 
sheet and is based on nonorthogonal basic vectors. The corresponding   
labelling, $(i,j,\rho)$, involves index 
$i$ numerating the sites around the nanotube,  spiral index
$j$ and index $\rho=0,1$ that describes sites 
that have the nearest neighbours  `down' ($\rho=0$) or `up' 
($\rho=1$), as shown in Fig.1. 

Note further that a hexagonal nanotube possesses two symmetries: the 
translation along the axis of the nanotube by $3d$ and the rotation 
by an angle $\alpha$ around the nanotube axis.  Given this, one can 
use an alternative labelling scheme in which the basic unit cell is 
rectangular and contains four atoms.  This scheme, also shown in Fig. 
1, involves using the labelling $(m,n,\varrho)$ where $m$ is the 
axial index, $n$ is the azimuthal index and the index 
$\varrho=1,2,3,4$ enumerates the atoms in the unit cell.

The position of any nanotube lattice site, at its equilibrium, can be 
described by $\vec{R}^0_{\ae}$ given by
\begin{equation}
\vec{R}^0_{\ae}
= R (\vec{e}_x \sin \Theta_{\ae}
 + \vec{e}_y \cos \Theta_{\ae} )
 + \vec{e}_z z_{\ae},
\label{Rnmj}
\end{equation}
where the three-component index  $\ae = \{\ae_1, \ae_2, \ae_3 \}$ 
indicates the nanotube atoms, and the coordinates $\Theta $, being an 
azimuthal angle and $z$, being a coordinate along the tube, describe 
positions of atoms on the cyllindrical surface of the nanotube.  
In the first scheme $\ae = \{i,j,\rho\}$ and in the second one $\ae = 
\{m,n,\varrho \}$.  In a zigzag nanotube the azimuthal and 
longitudinal positions of atoms are:  
\begin{eqnarray} 
\Theta_{i,j,\rho} = (i+{j+\rho\over 2})\alpha ;& \Theta_{m,n,1} = 
\Theta_{m,n,4} = n\alpha , & \Theta_{m,n,2} = \Theta_{m,n,3} = 
(n+\frac{1}{2})\alpha ; \nonumber\\ z_{i,j,\rho} = {3j+\rho\over 2}d 
;& z_{m,n,\varrho=1,2} = (3m-1+\frac{\varrho-1}{2})d ,& 
z_{m,n,\varrho=3,4} = (3m+1+\frac{\varrho-4}{2})d.
\label{Thetanmj4}
\end{eqnarray}

Although in the numerical work reported
in \cite{us}, we have used the first description, the second 
one is more convenient when taking into account the boundary 
conditions.  The azimuthal periodic condition $f(n+N)=f(n)$ is 
natural because going from $n$ to $n+N$ corresponds to a rotation by 
$2\pi$. In the $m$ direction, however, for a long enough nanotube, we 
can use the Born-Karman periodic conditions for electron  and phonon 
states in a nanotube. Thus, nanotubes can be considered as 1D systems 
with a complex inner structure.

Next, we consider displacements from the equilibrium positions of the 
all sites of the nanotube:
\begin{equation}
\vec{R}_{\ae} = \vec{R}^0_{\ae} + \vec{U}_{\ae},
\end{equation}
where the local displacement vector can be represented as the three 
orthogonal local vectors:
\begin{equation}
\vec{U}_{\ae} = \vec{u}_{\ae} + \vec{s}_{\ae}
                     + \vec{v}_{\ae}.
\end{equation}
Here $\vec{u}_{\ae}$ is tangent to the surface of the undeformed 
nanotube and perpendicular to the nanotube axis, $\vec{v}_{\ae}$ is 
tangent to this surface and parallel to the nanotube axis, and 
 $\vec{s}_{\ae}$ is normal to the surface of the nanotube. Then, 
using Cartesian coordinates, we have
\begin{eqnarray}
 \vec{u}_{\ae} &=&  u_{\ae}
      (\vec{e_x} \cos \Theta_{\ae}
    - \vec{e_y} \sin \Theta_{\ae} ) ,
\nonumber \\
\vec{s}_{\ae} &=&  s_{\ae} (\vec{e_x}  \sin \Theta_{\ae}   +
    \vec{e_y}  \cos \Theta_{\ae} ) ,
 \nonumber \\
 \vec{z}_{\ae} &=&  v_{\ae} \vec{e_z}.
\end{eqnarray}

To write down the Hamiltonian in a compact form, it is convenient to 
define the formal index operators of lattice translations: $r()$, 
$l()$ and $d()$,  which when applied to any lattice site index, 
translate the index to one of the three nearest sites. Applying these 
operators to the lattice site which has the nearest neighbour down, 
{\it i.e.} which in the first formulation have the index $\rho=0$, 
they translate the index respectively to the right, left and down 
from that site. For the lattice sites which have an upper nearest 
neighbour, {\it i.e.} which in the first formulation have the index 
$\rho=1$, one has to turn the lattice upside down before applying 
these definitions. Notice that the square of each of these three 
operators is equivalent to the identity operator. So, for example, 
moving from a lattice site to the right once and then moving to the 
right again, after flipping the lattice upside down, one returns to 
the starting site.
In particular, we have for the first lattice parametrisation
\begin{eqnarray}
r(i,j,0) = (i,j,1),  \qquad && r(i,j,1) = (i,j,0),\nonumber\\
l(i,j,0) = (i-1,j,1),\qquad && l(i,j,1) = (i+1,j,0),\nonumber\\
d(i,j,0) = (i,j-1,1),\qquad && d(i,j,1) = (i,j+1,0),
\label{indexop1}
\end{eqnarray}
while for the second one, which we will use below, we have
\begin{eqnarray}
r(m,n,1) = (m,n,2),  \qquad && r(m,n,2) = (m,n,1),\nonumber\\
r(m,n,3) = (m,n+1,4),\qquad && r(m,n,4) = (m,n-1,3),\nonumber\\
l(m,n,1) = (m,n-1,2),\qquad && l(m,n,2) = (m,n+1,1),\nonumber\\
l(m,n,3) = (m,n,4),  \qquad && l(m,n,4) = (m,n,3),  \nonumber\\
d(m,n,1) = (m-1,n,4),\qquad && d(m,n,2) = (m,n,3),\nonumber\\
d(m,n,3) = (m,n,2),  \qquad && d(m,n,4) = (m+1,n,1).
\label{indexop2}
\end{eqnarray}

Some physical quantites, e.g. the potential energy of the lattice distortion,
include central forces, which
depend on the distance between two sites.
Let us define the following lattice vectors connecting the atom 
$\{\ae\}$ with its three nearest neighbours $\delta(\ae)$ with 
$\delta = r,l,d$ for the right ($r$), left ($l$) and down or up ($d$) 
neighbours:
\begin{eqnarray}
\vec{D\delta}_{\ae} = \vec{R}_{\delta(\ae)}   - \vec{R}_{\ae}
 =\vec{D\delta}^0_{\ae}+ (\vec{U}_{\delta(\ae)} - \vec{U}_{\ae} ).
\end{eqnarray}
When $\vec{U}_{\ae}=0$  we add the upper index $0$ to all quantities 
 to indicate their values at the equilibrium position.  Note that 
$|\vec{Dr}^0_{\ae}| = |\vec{Dl}^0_{\ae}| = |\vec{Dd}^0_{\ae}|= d$.  
In the case of small displacements, {\it i.e} when $ 
|\vec{U}_{\delta(\ae)} - \vec{U}_{\ae}| \ll d$, the distance between 
the lattice sites is approximately given by:
\begin{equation}
|\vec{D\delta}_{\ae}| \approx d + W\delta_{\ae},
\end{equation}
where
\begin{equation}
 W\delta_{\ae} = \frac{(\vec{U}_{\delta(\ae)} - \vec{U}_{\ae} )
\cdot \vec{D\delta}^0_{\ae}}{d}
\end{equation}
are the changes of the distances between the nearest neighbours due 
to site displacements. The explicit expressions for $W\delta_{\ae}$ 
in the first scheme are 
\begin{eqnarray} 
&&Wr_{i,j,0} 
= {\sqrt{3}\over2}\Big( \cos(\frac{\alpha}{4})(u_{i,j,1}-u_{i,j,0})+ 
         \sin(\frac{\alpha}{4})(s_{i,j,1}+s_{i,j,0})\Big)
         + \frac{1}{2} (v_{i,j,1} - v_{i,j,0})\nonumber\\
&&Wl_{i,j,0} = {\sqrt{3}\over2}\Big(
        \cos(\frac{\alpha}{4})(u_{i,j,0}-u_{i-1,j,1})+
         \sin(\frac{\alpha}{4})(s_{i-1,j,1}+s_{i,j,0})\Big)
         + \frac{1}{2} (v_{i-1,j,1} - v_{i,j,0}) \nonumber\\
&&Wd_{i,j,0} = -v_{i,j-1,1} + v_{i,j,0}\nonumber\\
&&Wr_{i,j,1} = Wr_{i,j,0},\qquad Wl_{i,j,1} = Wl_{i+1,j,0},\qquad
Wd_{i,j,1} = Wd_{i,j+1,0}.
\end{eqnarray}

Because the central forces  between neighbouring sites do not provide
lattice stability, in addition to $W\delta_{\ae}$,
which are invariant under translations, we need also the
quantities $\Omega \delta_{\ae}$
which describe relative shifts of neighbouring sites. The  
corresponding explicit expressions are:
\begin{eqnarray}
&&\Omega r_{i,j,0} = {1\over2}\Big(
        \cos(\frac{\alpha}{4})(u_{i,j,1}-u_{i,j,0})+
         \sin(\frac{\alpha}{4})(s_{i,j,1}+s_{i,j,0})\Big)
         - \frac{\sqrt{3}}{2} (v_{i,j,1} - v_{i,j,0}),\nonumber\\
&&\Omega l_{i,j,0} = {1\over2}\Big(
        \cos(\frac{\alpha}{4})(u_{i,j,0}-u_{i-1,j,1})+
         \sin(\frac{\alpha}{4})(s_{i-1,j,1}+s_{i,j,0})\Big)
         - \frac{\sqrt{3}}{2} (v_{i-1,j,1} - v_{i,j,0}), \nonumber\\
&&\Omega d_{i,j,0} = -u_{i,j-1,1} + u_{i,j,0},\qquad
\Omega r_{i,j,1} = \Omega r_{i,j,0},\nonumber\\
&&\Omega l_{i,j,1} = \Omega l_{i+1,j,0},\qquad
\Omega d_{i,j,1} = \Omega d_{i,j+1,0}.
\end{eqnarray}

Note, the curvature of the lattice and corresponding bond-bending
in nanotubes is an important factor for the lattice stability \cite{WoMah}
and electron-phonon interaction \cite{Kempa}.
To take into account this factor we choose to based our discussion
on the solid angle spanned by the 3 lattice vectors at a given site:
\begin{eqnarray}
S_{\ae} = {(\vec{Dl}_{\ae}
  \times \vec{Dr}_{\ae}) \cdot  \vec{Dd}_{\ae}\over
     |\vec{Dr}_{\ae}| |\vec{Dl}_{\ae}| |\vec{Dd}_{\ae}|}
 \approx
 S^0_{\ae} + \frac{\sqrt{3}}{2d} C_{\ae}
\end{eqnarray}
where $S^0_{\ae} = \frac{3}{4} \sin(\frac{\alpha}{2})$ and,
in the case of small displacements,
\begin{eqnarray}
C_{i,j,0} &=& {\sqrt{3}\over 4}
          \sin(\frac{\alpha}{2}) (2 v_{i,j,0}-v_{i,j,1} - v_{i-1,j,1})
                                 \nonumber\\
&& - \cos(\frac{\alpha}{4}) s_{i,j-1,1} + 3\cos^3(\frac{\alpha}{4})s_{i,j,0}
\nonumber\\
&&  + (\frac{3}{2}\cos(\frac{\alpha}{4})-\frac{5}{2}\cos^3(\frac{\alpha}{4}))
      (s_{i-1,j,1}+s_{i,j,1})\nonumber\\
&& + \sin(\frac{\alpha}{4})(\frac{5}{2}\cos^2(\frac{\alpha}{4}) -1)
         (u_{i,j,1}-u_{i-1,j,1}),\nonumber\\
C_{i,j,1} &=& {\sqrt{3}\over 4}
          \sin(\frac{\alpha}{2}) (v_{i,j,0} + v_{i+1,j,0}- 2 v_{i,j,1})
                 \nonumber\\
&& - \cos(\frac{\alpha}{4}) s_{i,j+1,0}+3\cos^3(\frac{\alpha}{4})s_{i,j,1}
\nonumber\\
&&  + (\frac{3}{2}\cos(\frac{\alpha}{4})-\frac{5}{2}\cos^3(\frac{\alpha}{4}))
      (s_{i+1,j,0}+s_{i,j,0})\nonumber\\
&&  + \sin(\frac{\alpha}{4})(\frac{5}{2}\cos^2(\frac{\alpha}{4}) -1)
      (u_{i+1,j,0}-u_{i,j,0}).\nonumber\\
\end{eqnarray}
It is easy to write down the corresponding expressions in the second 
labelling scheme. This time one has twice the number of the 
expressions as compared with the first scheme.

\section{Phonon Hamiltonian}

We define the phonon Hamiltonian in the nearest-neighbour interaction 
approximation and take into account the potential terms responsible for the 
central, $V_{W}$, non-central, $V_{\Omega}$,  and the bond-bending,
$V_{C}$ forces in the harmonic approximation:
\begin{equation}
H_{ph} = \frac{1}{2} \sum_{\ae} \Bigl({{\vec P}_
{\ae}^2\over M}\ +\  k\sum_{\delta }
[W\delta_{\ae}^2 +\Omega \delta _{\ae}^2]\ +\ k_c C_{\ae} 
\Bigr),
\label{phon-ham1}
\end{equation}
where $M$ is the atom mass, $k$ is the  elasticity constant for the 
relative atom displacements, $k_c$ is a characteristic constant of 
the bond-bending force while ${\vec P}_{\ae}$ is the momentum, 
canonically conjugate to the displacement ${\vec U}_{\ae}$.

According to the theory of lattice dynamics (see, e.g., 
\cite{Maradudin}) the Hamiltonian (\ref{phon-ham1}) can be 
diagonalised by some unitary transformation. For the lattice 
labelling $\ae =\{m,n,\varrho\}$, this transformation has the form
\begin{eqnarray}
u_{m,n,\varrho} = \frac{1}{\sqrt{12MNL}} \sum_{k,\nu,\tau}e^{i(km+\nu
n)}U_{\varrho,\tau}(k,\nu)Q_{k,\nu,\tau},
\nonumber\\
s_{m,n,\varrho} = \frac{1}{\sqrt{12MNL}} \sum_{k,\nu,\tau}e^{i(km+\nu
n)}S_{\varrho,\tau}(k,\nu)Q_{k,\nu,\tau},
\nonumber\\
v_{m,n,\varrho} = \frac{1}{\sqrt{12MNL}} \sum_{k,\nu,\tau}e^{i(km+\nu
n)}V_{\varrho,\tau}(k,\nu)Q_{k,\nu,\tau}.
\label{phtransf}
\end{eqnarray}

Then, introducing the operators of creation,
$b\sp{\ast}_{k,\nu,\tau}$, and annihilation, $b_{k,\nu,\tau}$, of phonons
\begin{equation}
Q_{k,\nu,\tau}\,=\,\sqrt{\frac{\hbar}{2\omega_{\tau}(k,\nu)}}
\left(b_{k,\nu,\tau}\,+\,b_{-k,-\nu,\tau}\sp{\dagger}\right),
\label{ncoor}
\end{equation}
we can rewrite the phonon Hamiltonian (\ref{phon-ham1}) in the standard form
\begin{eqnarray}
H_{ph}&=&\, \frac{1}{2} \sum_{k,\nu,\tau} \Bigl(
P_{k,\nu,\tau}\sp{\dagger}P_{k,\nu,\tau} + \omega^2_{\tau}(k,\nu)
Q_{k,\nu,\tau}\sp{\dagger} Q_{k,\nu,\tau} \Bigr)\nonumber\\
&=&\,\sum_{q,\nu \tau}\, \hbar 
\omega_{\tau}(q,\nu)\left(b_{q,\nu,\tau}\sp{\dagger}b_{q,\nu,\tau} 
+ \ {1\over 2}\right).
\label{omega}
\end{eqnarray}
Here $\omega_{\tau}(k,\nu)$ is the frequency of the normal lattice vibrations
of the mode $\tau$ ($\tau = 1,2,\dots,12$) with the longitudinal
wavenumber $k$ and the azimuthal quantum number $\nu$. The adimensional 
wavenumber (quasi-momentum) along the nanotube, $k = \frac{2\pi}{L}n_1$,
takes quasi-continuum values (for $L \gg 1$) in the
range $-\pi < k \leq \pi$. The azimuthal quantum number takes
discrete values $\nu = \frac{2\pi}{N}n_2$ with $n_2=0,\pm
1,\dots,\pm \frac{N-1}{2}$ if $N$ is odd and $n_2=0,\pm 1,\dots,\pm
(\frac{N}{2}-1),\frac{N}{2}$ if $N$ is even.

The frequencies $\omega_{\tau}(k,\nu)$ and the coefficients of the 
transformation (\ref{phtransf}) can be found from the diagonalization 
condition of the potential energy of the lattice displacements in 
(\ref{phon-ham1}) with the orthonormalization conditions 
$$\frac{1}{12}\sum_{\varrho}\left( U_{\varrho,\tau}\sp{\ast}(k,\nu) 
U_{\varrho,\tau'}(k,\nu)+ S_{\varrho,\tau}\sp{\ast}(k,\nu) 
S_{\varrho,\tau'}(k,\nu) + V_{\varrho,\tau}\sp{\ast}(k,\nu) 
V_{\varrho,\tau'}(k,\nu) \right) = \delta_{\tau,\tau'},$$
$$\frac{1}{12}\sum_{\tau} U_{\varrho,\tau}\sp{\ast}(k,\nu) 
U_{\varrho',\tau}(k,\nu)= \frac{1}{12}\sum_{\tau} 
S_{\varrho,\tau}\sp{\ast}(k,\nu) S_{\varrho',\tau}(k,\nu) = 
\frac{1}{12}\sum_{\tau} V_{\varrho,\tau}\sp{\ast}(k,\nu) V_{\varrho',\tau}(k,\nu) =
\delta_{\varrho,\varrho'},$$
\begin{equation}
\sum_{\tau} U_{\varrho,\tau}\sp{\ast}(k,\nu) S_{\varrho',\tau}(k,\nu)=
\sum_{\tau} S_{\varrho,\tau}\sp{\ast}(k,\nu) V_{\varrho',\tau}(k,\nu) =
\sum_{\tau} V_{\varrho,\tau}\sp{\ast}(k,\nu) U_{\varrho',\tau}(k,\nu) = 0.
\label{phortonorm2}
\end{equation} 

Note that any linear form of lattice displacements, such as 
$W\delta_{m,n,\varrho}$, $\Omega \delta_{m,n,\varrho}$ and 
$C_{m,n,\varrho}$, after applying the transformation (\ref{phtransf}) 
can be written as
\begin{equation}
F_{m,n,\varrho}= \frac{1}{\sqrt{12MNL}} \sum_{k,\nu,\tau}e^{i(km+\nu n)}
F_{\varrho}(k,\nu|\tau)Q_{k,\nu,\tau}
\end{equation} 
where $F_{\varrho}(k,\nu|\tau)$ is a linear form of the 
transformation coefficients $S_{\varrho,\tau}(k,\nu)$, $ 
V_{\varrho,\tau}(k,\nu)$ and $U_{\varrho,\tau}(k,\nu)$. 

Therefore, in general, the frequencies $\omega_{\tau}(k,\nu)$ of the normal 
vibrations of the lattice can be represented as
\begin{equation}
\omega_{\tau}^2(k,\nu) = \frac{1}{12}\sum_{\varrho} \Bigl(
 \frac{k}{M}\sum_{\delta }
 [|W\delta _{\varrho}(k,\nu|\tau)|^2
+|\Omega \delta _{\varrho}(k,\nu|\tau)|^2] +
\frac{k_c}{M} |C_{\varrho}(k,\nu|\tau)|^2\Bigr).
\label{omega1}
\end{equation}

The equations for the normal modes of lattice vibrations are too
complicated to be solved analytically in the general case. For 
carbon nanotubes the phonon modes were calculated numerically (see, e.g., 
\cite{SaiDrDr,Mah_vibr} and references therein). Here we do not 
calculate the phonon spectrum explicitely, instead we use the general 
relations (\ref{phortonorm2}),(\ref{omega1}) to get some estimates 
which only depend on the parameters $k$, $k_c$ and $M$.  Meanwhile, 
the explicit expressions for the electron dispersions are more 
important for us and will be derived below.

\section{Electron Hamiltonian}

The electron eigenstates are found from the tight-binding model using 
the nearest-neighbour hopping approximation. In this approximation 
the Hamiltonian which describes electron states is given by 
\begin{eqnarray}
H_e &=&\,\sum_{\ae,\sigma} \Bigl({\cal 
E}_0\,a_{\ae,\sigma}\sp{\dagger}a_{\ae,\sigma}\, - 
J\,\sum_{\delta} a_{\ae,\sigma}\sp{\dagger}
a_{\delta(\ae),\sigma} \Bigr).
\label{Hamilt1_mn}
\end{eqnarray}
Here $a_{\ae,\sigma}\sp{\dagger}$($ a_{\ae,\sigma}$) 
are creation (annihilation) operators of a $\pi$-electron with 
the spin $\sigma$ on the site $\ae$, ${\cal E}_0$ is the 
$\pi$-electron energy, $J$ is the energy of
 the hopping interaction between the nearest 
neighbours and the summation over $\delta$ denotes the summation over 
the three nearest neighbour sites.

By the unitary transformation
\begin{equation}
a_{m,n,\varrho,\sigma} = \frac{1}{2\sqrt{LN}} 
\sum_{k,\nu,\lambda}e^{ikm+i\nu n} 
u_{\varrho,\lambda}(k,\nu)c_{k,\nu,\lambda ,\sigma},
\label{etransf}
\end{equation}
with
\begin{equation}
\frac{1}{4}\sum_{\varrho}u_{\varrho,\lambda}\sp{\ast}(k,\nu) 
u_{\varrho,\lambda'}(k,\nu)= \delta_{\lambda,\lambda'}
\end{equation}
the Hamiltonian (\ref{Hamilt1_mn}) is transformed into a diagonal form 
(see Appendix 1):
\begin{equation}
H_e =\,\sum_{k,\nu,\lambda,\sigma} E_{\lambda}(k,\nu)\,
c_{k,\nu,\lambda,\sigma}\sp{\dagger}c_{k,\nu,\lambda,\sigma}\,.
\label{Hamilt2}
\end{equation}
Here $k$ is an adimensional quasi-momentum along the 
nanotube, $\nu$ is an azimuthal quantum number, and  $\lambda = 1,2,3,4$ 
labels the four series (due to the four atoms in each cell), of 
1D electronic bands with the dispersion laws
\begin{equation}
E_{\lambda}(k,\nu)\,=\,{\cal E}_0\,\pm\,{\cal E}_{\pm}(k,\nu),
\label{bands}
\end{equation}
where
\begin{equation}
{\cal E}_{\pm}(k,\nu)\,=\,J\,\sqrt{1+4\cos^2(\frac{\nu}{2}) \pm 
4\cos(\frac{\nu}{2})\cos(\frac{k}{2})}\,.
\label{Epm}
\end{equation}
In (\ref{Hamilt2}) the operators $c_{k,\nu,\lambda,\sigma}\sp{\dagger}$($ 
c_{k,\nu,\lambda,\sigma}\,$) are creation (annihilation) operators of 
electrons with the quasi-momentum $k$ and spin $\sigma$ in the band 
($\nu,\lambda$). If we label the electronic bands as
\begin{eqnarray}
&& E_{1}(k,\nu)\,=\,{\cal E}_0\,-\,{\cal E}_{+}(k,\nu),\qquad
E_{2}(k,\nu)\,=\,{\cal E}_0\,-\,{\cal E}_{-}(k,\nu),\nonumber\\
&&E_{3}(k,\nu)\,=\,{\cal E}_0\,+\,{\cal E}_{-}(k,\nu),\qquad
E_{4}(k,\nu)\,=\,{\cal E}_0\,+\,{\cal E}_{+}(k,\nu),
\label{E1234}
\end{eqnarray}
then the matrix of the unitary transformation coefficients 
${\bf u}$ (\ref{etransf}) is given by
\begin{equation}
{\bf u}(k,\nu)=\left( \begin{array}{cccc}
e^{-i(\frac{k+\nu}{4}+\theta_{+})} & e^{-i(\frac{k+\nu}{4}-\theta_{-})} &
e^{-i(\frac{k+\nu}{4}-\theta_{-})} & e^{-i(\frac{k+\nu}{4}+\theta_{+})} \\
e^{-i(\frac{k-\nu}{4}-\theta_{+})} & e^{-i(\frac{k-\nu}{4}+\theta_{-})} &
-e^{-i(\frac{k-\nu}{4}+\theta_{-})} & -e^{-i(\frac{k-\nu}{4}-\theta_{+})} \\
e^{i(\frac{k+\nu}{4}-\theta_{+})} & -e^{i(\frac{k+\nu}{4}+\theta_{-})} &
-e^{i(\frac{k+\nu}{4}+\theta_{-})} & e^{i(\frac{k+\nu}{4}-\theta_{+})} \\
e^{i(\frac{k-\nu}{4}+\theta_{+})} & -e^{i(\frac{k-\nu}{4}-\theta_{-})} &
e^{i(\frac{k-\nu}{4}-\theta_{-})} & -e^{i(\frac{k-\nu}{4}+\theta_{+})}
\end{array}
\right),
\label{etr-coef}
\end{equation}
where the phases $\theta $ satisfy the 
relation (\ref{theta}), given in Appendix 1.

\section{Electron-Phonon Hamiltonian}

The electron-phonon interaction originates from different mechanisms 
\cite{MDWh,JiDrDr,Kane,WoMah,Mah}.  Usually, 
the dependence of the hopping interaction between the 
nearest neighbours $J_{(\ae);\delta(\ae)}$ on their separation 
is considered and in the linear approximation with respect to the 
displacements one has 
\begin{equation}
 J_{(\ae);\delta(\ae)} = J - G_2 W\delta_{\ae}. 
\end{equation} 
In general, neighbouring atoms also 
alter the energy of the $\pi$-electrons on each site and so, in the same 
linear approximation, we can write 
\begin{equation}
{\cal E}_{\ae} = {\cal 
E}_0 + \chi_1\sum_{\delta }W \delta_ \ae +\chi_2\,C_{\ae}\,.
\end{equation}

Thus, the total electron-phonon interaction Hamiltonian should be 
taken in the following form 
\begin{equation}
H_{int} =\,\sum_{\ae,\sigma} 
\Bigl(\,a_{\ae,\sigma}\sp{\dagger}a_{\ae,\sigma}\,
[\chi_1\,\sum_{\delta}W\delta_{\ae}
 +\chi_2\,C_{\ae}]+
G_2\,\sum_{\delta }a_{\ae,\sigma}\sp{\dagger}a_{\delta (\ae),\sigma } 
W\delta _\ae \,\Bigr),
\label{Hint1}
\end{equation}
where we have used the translation index operator $\delta (\ae)$
defined in (\ref{indexop2}).

The unitary transformations (\ref{etransf}) and 
(\ref{phtransf}), transform the interaction Hamiltonian into 
\begin{equation}
H_{int} 
=\frac{1}{2\sqrt{3LN}}\sum_{k,\nu,\lambda,\lambda',q,\mu,\tau,\sigma}
F_{\lambda,\lambda'}^{(\tau)}(k,\nu;q,\mu)
c_{k+q,\nu +\mu,\lambda',\sigma}\sp{\dagger}c_{k,\nu,\lambda,\sigma}
Q_{q,\mu,\tau}
\label{Hint2}
\end{equation}
where $Q_{q,\mu,\tau}$ was determined in (\ref{ncoor}) and 
\begin{equation}
F_{\lambda',\lambda}^{(\tau)}(k,\nu;q,\mu) = 
\frac{1}{4}\sum_{\varrho',\varrho} u_{\varrho',\lambda'}(k+q,\nu + 
\mu)\sp{\ast} T_{\varrho',\varrho}(k,\nu;q,\mu|\tau) 
u_{\varrho,\lambda}(k,\nu).
\label{F}
\end{equation}

Note that $T_{\varrho',\varrho}(k,\nu;q,\mu|\tau) = 
T_{\varrho,\varrho'}\sp {\ast}(k+q,\nu+\mu;-q,-\mu|\tau)$ and that 
 $T_{1,3}=T_{3,1}=T_{2,4}=T_{4,2}=0$.  The diagonal elements, at 
$\varrho'=\varrho$, are
\begin{equation}
T_{\varrho,\varrho}(q,\mu|\tau) = \frac{\chi_1}{\sqrt{M}} 
\,W_{\varrho}(q,\mu|\tau) + \frac{\chi_2}{\sqrt{M}} \,C_\varrho 
(q,\mu|\tau),
\label{H-jj}
\end{equation}
and the nonzero off-diagonal elements, $\varrho \neq \varrho'$, are given by
\begin{equation}
T_{\varrho',\varrho}(k,\nu;q,\mu|\tau) = \frac{G_2}{\sqrt{M}} 
\,W_{\varrho',\varrho}(k,\nu;q,\mu|\tau),
\label{H-jj'}
\end{equation}
where $W_{\varrho}(q,\mu|\tau)$, $C_\varrho (q,\mu|\tau)$ and 
$W_{\varrho',\varrho}(k,\nu;q,\mu|\tau)$ are determined only by the 
coefficients of the phonon unitary transformation (\ref{phtransf}). 
In particular, \begin{eqnarray} W_{1}(q,\mu|\tau) &=& 
\sqrt{3}\sin(\frac{\alpha}{4})\left(S_{1,\tau} + 
e^{-i\frac{\mu}{2}}\cos(\frac{\mu}{2})S_{2,\tau})\right)+ \nonumber\\ 
&+& 
i\sqrt{3}\cos(\frac{\alpha}{4})\sin(\frac{\mu}{2})e^{-i\frac{\mu}{2}} 
U_{2,\tau}+ \cos(\frac{\mu}{2})e^{-i\frac{\mu}{2}} V_{2,\tau} - 
e^{-iq}V_{4,\tau}, \nonumber\\ W_{2}(q,\mu|\tau) &=& 
\sqrt{3}\sin(\frac{\alpha}{4})\left(S_{2,\tau} + 
\cos(\frac{\mu}{2})e^{i\frac{\mu}{2}}S_{1,\tau})\right) + \nonumber\\ 
&+& 
i\sqrt{3}\cos(\frac{\alpha}{4})\sin(\frac{\mu}{2})e^{i\frac{\mu}{2}}U_{1,\tau} 
- \cos(\frac{\mu}{2})e^{i\frac{\mu}{2}} V_{1,\tau} + V_{3,\tau},
\nonumber\\
W_{3}(q,\mu|\tau) &=& \sqrt{3}\sin(\frac{\alpha}{4})\left(S_{3,\tau} +
\cos(\frac{\mu}{2})e^{i\frac{\mu}{2}}S_{4,\tau})\right) +
\nonumber\\
&+&
i\sqrt{3}\cos(\frac{\alpha}{4})\sin(\frac{\mu}{2})e^{i\frac{\mu}{2}}U_{4,\tau}+
\cos(\frac{\mu}{2})e^{i\frac{\mu}{2}} V_{4,\tau} - V_{2,\tau},
\nonumber\\
W_{4}(q,\mu|\tau) &=& \sqrt{3}\sin(\frac{\alpha}{4})\left(S_{4,\tau} +
\cos(\frac{\mu}{2})e^{-i\frac{\mu}{2}}S_{3,\tau})\right) +
\nonumber\\
&+&
i\sqrt{3}\cos(\frac{\alpha}{4})\sin(\frac{\mu}{2})e^{-i\frac{\mu}{2}}U_{3,\tau}
-\cos(\frac{\mu}{2})e^{-i\frac{\mu}{2}} V_{3,\tau} + e^{iq}V_{1,\tau},
\label{W-j}
\end{eqnarray}
and
\begin{eqnarray}
W_{12}(\nu;q,\mu|\tau) &=& e^{-i\frac{\nu}{2}}
  \Bigl(\sqrt{3}\sin(\frac{\alpha}{4})\left(
\cos(\frac{\nu}{2})S_{1,\tau}+ e^{-i\frac{\mu}{2}}
\cos(\frac{\nu+\mu}{2})S_{2,\tau})\right) -
\nonumber\\
&-&
i\sqrt{3}\cos(\frac{\alpha}{4})\left(\sin(\frac{\nu}{2})U_{1,\tau} -
e^{-i\frac{\mu}{2}}\sin(\frac{\nu+\mu}{2})U_{2,\tau})\right)+
\nonumber\\
&+&
e^{-i\frac{\mu}{2}}
\cos(\frac{\nu+\mu}{2}) V_{2,\tau} - \cos(\frac{\nu}{2})V_{1,\tau}\Bigr),
\nonumber\\
W_{14}(k;q,\mu|\tau) &=& e^{-ik}\left( V_{1,\tau} - e^{-iq}V_{4,\tau} \right), 
\qquad 
W_{23}(q,\mu|\tau) =  V_{3,\tau} - V_{2,\tau}, \nonumber \\
W_{34}(\nu;q,\mu|\tau) &=& e^{i\frac{\nu}{2}}
  \Bigl(\sqrt{3}\sin(\frac{\alpha}{4})\left(
\cos(\frac{\nu}{2})S_{3,\tau}+ e^{i\frac{\mu}{2}}
\cos(\frac{\nu+\mu}{2})S_{4,\tau})\right) -
\nonumber\\
&-&
i\sqrt{3}\cos(\frac{\alpha}{4})\left(\sin(\frac{\nu}{2})U_{3,\tau} -
e^{i\frac{\mu}{2}}\sin(\frac{\nu+\mu}{2})U_{4,\tau})\right)+
\nonumber\\
&+&
e^{i\frac{\mu}{2}}
\cos(\frac{\nu+\mu}{2}) V_{4,\tau} - \cos(\frac{\nu}{2})V_{3,\tau}\Bigr),
\label{W-jj'}
\end{eqnarray}
where $S_{\varrho,\tau}=S_{\varrho,\tau}(q,\mu)$, 
$V_{\varrho,\tau}=V_{\varrho,\tau}(q,\mu)$, and 
$U_{\varrho,\tau}=U_{\varrho,\tau}(q,\mu)$.

Thus, the functions $F_{\lambda,\lambda'}^{(\tau)}(k,q;\nu,\mu)$ are 
determined by 
the interaction parameters $\chi_1$, $\chi_2$, $G_2$ and the coefficients 
of the unitary transformations (\ref{etransf}) and (\ref{phtransf}).  
Note that, in (\ref{Hint2}), the azimuthal numbers satisfy the relation 
$\nu_1 = \nu +\mu$, for which the following rule should be applied: if 
$|\nu +\mu| > 2\pi$, then $\nu_1 \rightarrow \nu'_1 = \nu_1 \pm 2\pi$ 
in such a way that $|\nu'_1| < 2\pi$.

\section{Adiabatic approximation}

The total Hamiltonian of the system is then given by
\begin{equation}
H\,=\,H_e\,+\,H_{ph}\,+\,H_{int},
\label{Htot}
\end{equation}
where $H_e$, $H_{ph}$ and $H_{int}$ are given by 
(\ref{Hamilt2}), (\ref{omega}) and (\ref{Hint2}), respectively.

Below we consider only one-particle states in a carbon nanotube 
taking into account the interaction of the particle with the lattice 
distortions. When the coupling constant of this interaction is strong 
enough, this interaction can lead to the self-trapping of the 
particle.  The self-trapped states are usually  described in the 
adiabatic approximation.  In this approximation the wavefunction of 
the system is represented as
\be
|\Psi\rangle = U\,|\psi_e \rangle, 
\label{adappr}
\ee
where $U$ is a unitary operator of the coherent atom displacements 
induced by the presence of the quasiparticle and so is given by an 
expression of the form 
\be
U\ =\exp{\left [\sum_{\mu,q,\tau} 
(\beta_{\tau}(q,\mu)b\sp{\dagger}_{q,\mu,\tau}\ -\ 
\beta\sp{\ast}_{\tau}(q,\mu)b_{q,\mu,\tau})\right ]}
\label{uoperat}
\ee
and $|\psi_e \rangle $ is the wavefunction of the quasiparticle itself. 
Moreover we require that it satisfies
$\langle\psi_e |\psi_e \rangle = 1$.

In (\ref{uoperat}) the coefficients $\beta_{\mu,\tau}(q)$ depend on 
the state of the quasiparticle which, in turn, is determined by the 
lattice configuration. Using (\ref{adappr}) in the Schr\"{o}dinger 
equation $H|\Psi \rangle = E |\Psi \rangle$, we find the equation for 
the electronic part $|\psi_e \rangle $ of the total function 
(\ref{adappr})
\begin{equation}
\tilde{H}|\psi_e\rangle\,=\,E|\psi_e\rangle,
\label{eqtilde}
\end{equation}
where
\begin{equation}
\tilde{H}\,=\,U\sp{\dagger} H U\,=\, 
W\,+\,\tilde{H}_e\,+\,H_{int}\,+\,H_{ph}\,+\,H_d .
\label{Htilde}
\end{equation}
Here
\begin{equation}
W\,=\,\sum_{q,\mu,\tau}\, \hbar \omega_{\tau}(q,\mu)|\beta_{\tau}(q,\mu)|^2
\label{Wdef}
\end{equation}
is the energy of the lattice deformation,
\begin{eqnarray}
&&\tilde{H}_e = \sum_{k,\nu,\lambda,\sigma} E_{\lambda}(k,\nu)\,
c_{k,\nu,\lambda,\sigma}\sp{\dagger}c_{k,\nu,\lambda,\sigma}\,+\nonumber \\
&&+\frac{1}{2\sqrt{3LN}}\sum_{k,q,\lambda,\lambda',\tau,\sigma}
F_{\lambda',\lambda}^{(\tau)}(k,q;\nu,\mu)\,Q_{\tau}(q,\mu)
c_{k+q,\nu +\mu,\lambda',\sigma}\sp{\dagger}c_{k,\nu,\lambda,\sigma}
\label{Heltilda}
\end{eqnarray}
is the Hamiltonian of quasiparticles in the deformed lattice with 
 the deformation potential given by
\begin{equation}
Q_{\tau}(q,\mu)\,=\,\left(\frac{\hbar}{2\omega_{\tau}(q,\mu)} 
\right)\sp{\frac{1}{2}}
\left(\beta _{\tau}(q,\mu)\,+\,\beta\sp{\ast}_{\tau}(-q,-\mu)\right),
\label{Q-beta}
\end{equation}
and
\begin{equation}
H_d \,=\,\sum_{q,\mu,\tau}\, \hbar \omega_{\tau}(q,\mu)
(\beta_{\tau}(q,\mu)b\sp{\dagger}_{q,\mu,\tau}\ +\
\beta\sp{\ast}_{\tau}(q,\mu)b_{q,\mu,\tau})
\label{HL}
\end{equation}
is the displacement term of the phonon Hamiltonian. The latter term, 
$H_d$, is linear with respect to the phonon operators and appears 
here as a result of the action of the unitary operator 
(\ref{uoperat}).

With the help of the unitary transformation
\begin{equation}
c_{k,\nu,\lambda,\sigma}\,=\,\sum_{\eta}\psi_{\eta;\lambda }(k,\nu)
                                         C_{\eta,\sigma}, 
\label{c-C transf} 
\end{equation} 
we can introduce the new Fermi operators $C_{\eta,\sigma}$ for which, in the 
general case, the quantum number $\eta$ is a multicomponent index. 
The coefficients $ \psi_{\eta;\lambda} (k,\nu)\ $ are to be chosen from the 
condition that the electron Hamiltonian (\ref{Heltilda}) can be  
transformed into a diagonal form:
\begin{equation}
\tilde{H}_e\,=\,\sum_{\eta,\sigma} E_\eta 
  C_{\eta,\sigma}\sp{\dagger}C_{\eta,\sigma} .
\label{Hetildtr}
\end{equation}
This requirement leads to the following equations for the 
transformation coefficients:
\begin{eqnarray}
E_\eta \psi_{\eta;\lambda }(k,\nu) &=& 
E_{\lambda}(k,\nu)\psi_{\eta;\lambda }(k,\nu)+\nonumber\\
&+&\frac{1}{2\sqrt{3LN}}\sum_{q,\lambda',\tau}
F_{\lambda,\lambda'}^{(\tau)}(k-q,q;\nu-\mu,\mu)Q_{\tau}(q,\mu) 
\psi_{\eta;\lambda' }(k-q,\nu-\mu).
\label{eqpsi_j}
\end{eqnarray}
Solutions of this system of equations, with the orthonormalization 
condition
\be
\sum_{\lambda,\nu,k} \psi_{\eta;\lambda }\sp{\ast}(k,\nu) 
\psi_{\eta';\lambda }(k,\nu)=\delta_{\eta,\eta'}
\label{ortnorm}
\ee
then give us the coefficients $\psi_{\eta;\lambda }(k,\nu)$ as well as 
the eigenvalues $E_\eta$ of the electron energy levels.

After the transformation (\ref{c-C transf}) the interaction Hamiltonian 
becomes 
\begin{eqnarray} 
H_{int} 
&=&\,\frac{1}{2\sqrt{3LN}}\sum_{\eta,\eta',q,\mu,\tau,\sigma} 
\Gamma_{\eta,\eta'}^{(\tau)}(q,\mu)\,
C_{\eta,\sigma}\sp{\dagger}C_{\eta',\sigma}\, Q
_{q,\mu,\tau},
\label{Hint-transf}
\end{eqnarray}
where
\begin{eqnarray}
\Gamma_{\eta,\eta'}^{(\tau)}(q,\mu)\,&=& \sum_{k,\nu,\lambda,\lambda'}
 \psi_{\eta;\lambda'}\sp{\ast}(k+q,\nu+\mu)F_{\lambda', 
 \lambda}^{(\tau)}(k,q;\nu,\mu) \psi_{\eta';\lambda}(k,\nu).
\label{Phi_j}
\end{eqnarray}

The operator $H_{int}$ can be separated into two parts. The most important 
term, $H_{ad}$, is the diagonal part of $H_{int}$ with respect to the 
electron quantum 
numbers $\eta$ ($\eta=\eta'$ in (\ref{Hint-transf})). The remainder, 
$H_{na}$, the off-diagonal part of $H_{int}$, corresponds to 
phonon induced transitions between the adiabatic terms determined by Eqs. 
(\ref{eqpsi_j}). So we can represent the Hamiltonian (\ref{Htilde}) 
in the form 
\begin{equation} 
\tilde{H}\,=\,H_0\,+\,H_{na} 
\label{Hna}
\end{equation} 
where
\begin{equation}
H_0\,=\, W\,+\,\tilde{H}_e\,+\,H_{ad}\,+\,H_{ph}\,+\,H_d 
\label{H0}
\end{equation}
describes the system in the adiabatic approximation and 
$H_{na}$ is the nonadiabaticity operator.

At large enough electron-phonon 
coupling, the nonadiabaticity is less important and the operator 
$H_{na}$ can be considered as a perturbation. In the zero-order adiabatic  
approximation the quasiparticle wavefunction  $|\psi_e^{(0)} \rangle 
$ does not depend on phonon variables. In the case of a system 
with $N_e$ electrons it can be represented as a product of $N_e$ 
electron creation operators which act on the quasiparticle vacuum 
state. 

In particular, the one-particle states are described by the function
\be
|\psi_e^{(0)} \rangle =C_{\eta,\sigma}\sp{\dagger}|0\rangle,
\label{psi_e,gr1}
\ee
where $|0\rangle$ is the quasiparticle and phonons' vacuum state,  
the index $\eta$ labels the adiabatic state which is occupied by the 
quasiparticle. For the ground state we put $\eta=g$.  
The total wavefunction of 
the system (\ref{adappr}) describes the self-trapped states of a large 
polaron in the zero-order adiabatic approximation.

Note that the function (\ref{psi_e,gr1}) is an eigenstate of the 
zero-order adiabatic Hamiltonian $H_0$:  
\begin{eqnarray} 
H_0|\psi_e^{(0)} \rangle\,&=&\,\Bigl[W\,+\,E_g \,\nonumber\\ &+& 
\sum_{q,\mu,\tau}\Bigl( \bigl(\hbar \omega_{\tau}(q,\mu) \beta 
_{\mu,\tau}(q) +
  \frac{1}{2}\sqrt{\frac{\hbar}{6LN\omega_{\tau}(q,\mu)}}
  \Gamma_{g,g}^{(\tau)*}(q,\mu)\bigr)b
\sp{\dagger}_{q,\mu,\tau} +  h.c.\Bigr) \Bigr] |\psi_e^{(0)} \rangle
\label{adiabateq}
\end{eqnarray}
with the energy ${\cal E}_g = W + E_g$ provided that the coefficients
$\beta_{\tau}(q,\mu)$ in (\ref{uoperat}) satisfy:
\begin{eqnarray}
&&\hbar \omega_{\tau}(q,\mu) \beta _{\mu,\tau}(q) = 
 - \frac{1}{2}\sqrt{\frac{\hbar}{6LN\omega_{\tau}(q,\mu)}}
   \Gamma_{g,g}^{(\tau)*}(q,\mu) = \nonumber \\
&=& - \frac{1}{2}\sqrt{\frac{\hbar}{6LN\omega_{\tau}(q,\mu)}} 
   \sum_{k,\nu,\lambda,\lambda'}
      F_{\lambda,\lambda'}^{(\tau)*}(k,q;\nu,\mu)\psi_{g;\lambda'}
      \sp{\ast}(k,\nu) \psi_{g;\lambda}(k+q,\nu+\mu).
\label{eqbeta}
\end{eqnarray}

The adiabatic electron states are determined by (\ref{eqpsi_j}) in which the 
lattice distortion $Q_{\tau}(q,\mu)$, according to 
Eqs.(\ref{Q-beta},\ref{eqbeta}), is self-consistently determined by 
the electron state: 
\begin{equation}
Q_{\tau}(q,\mu)=
- \frac{1}{2\sqrt{3LN}}\sum_{k,\nu,\lambda,\lambda'} 
\frac{F_{\lambda,\lambda'}^{(\tau)*}(k,q;\nu,\mu)} 
{\omega_{\tau}^{2}(q,\mu)}\psi_{g;\lambda'}\sp{\ast}(k,\nu) 
\psi_{g;\lambda}(k+q,\nu+\mu).
\label{Q-psi}
\end{equation}
Substituting (\ref{Q-psi}) into equations 
(\ref{eqpsi_j}) for the occupied electron state, we obtain a 
nonlinear equation for $\psi_{g;\lambda}(k,\nu)$ whose  
solution, satisfying the normalization condition 
(\ref{ortnorm}), gives the wavefunction and eigenenergy $E_g$ of the 
electron ground state and, therefore, the self-consistent lattice 
distortion. All other unoccupied excited electron states with $\eta \neq 
g$ can be found from the linear equations (\ref{eqpsi_j}) with 
the given deformational potential.

Using the inverse unitary transformations (\ref{c-C transf}) and
(\ref{etransf}), we can rewrite the eigenfunction (\ref{psi_e,gr1}) in
the following form:
\be
|\psi_e^{(0)} \rangle =
\sum_{\lambda,\nu,k} \psi_{g;\lambda }(k,\nu)c_{k,\nu,\lambda,\sigma}
\sp{\dagger}|0\rangle =
\sum_{\ae} 
\psi_{g,\ae}a_{\ae,\sigma}\sp{\dagger}|0\rangle, 
\label{psi_e2}
\ee
where
\be
\psi_{g,\ae}=\frac{1}{2\sqrt{LN}} 
\sum_{\lambda,\nu,k} e^{i(km+\nu n)} u_{\varrho,\lambda}(k,\nu) 
\psi_{g;\lambda }(k,\nu).  
\label{psi_mnj} 
\ee 
Here $\psi_{g,\ae}$ is the polaron wave function, i.e.,  
the probability amplitude  of the distribution of a quasiparticle over 
the nanotube sites: $P(\ae) = |\psi_{g,\ae }|^2$.

\section{Large polaron state}

Putting Eqs (\ref{Q-psi}) into
(\ref{eqpsi_j}) gives us the nonlinear equations 
\begin{eqnarray}
\,\Bigl(E_{\lambda}(k,\nu)-E\Bigr)\psi_{\lambda }(k,\nu) =\, 
\nonumber\\
\frac{1}{LN}\sum_{\lambda',\lambda_1',\lambda_1,k_1,\nu_1,q,\mu}
G_{\lambda,\lambda'}^{\lambda_1',\lambda_1}\left( 
\begin{array}{ccc}k,&k_1,&q \\ 
\nu,&\nu_1,&\mu
\end{array} 
\right)
\psi_{\lambda_1}\sp{\ast}(k_1,\nu_1) \psi_{\lambda_1'}(k_1+q,\nu_1+ 
\mu)\psi_{\lambda' }(k-q,\nu-\mu)
\label{nleqpsi_g}
\end{eqnarray}
for the  one-electron ground state. Here, and from now onwards, we have 
omitted the index $\eta=g$ 
and introduced the notation 
\begin{equation} 
G_{\lambda,\lambda'}^{\lambda_1',\lambda_1}\left( 
\begin{array}{ccc}k,&k_1,&q \\ \nu,&\nu_1,&\mu
\end{array} 
\right)  =
\frac{1}{12}\sum_{\tau} 
\frac{F_{\lambda,\lambda'}^{(\tau)}(k-q,\nu-\mu;q,\mu) 
F_{\lambda_1',\lambda_1}^{(\tau)*}(k_1,\nu_1;q,\mu)}
{\omega_{\tau}^{2}(q,\mu)}.
\label{Gfunc}
\end{equation}

We see that all sub-levels of all sub-bands participate in the formation of 
the self-trapped electron states and, in general, there are many 
solutions of Eq.(\ref{nleqpsi_g}).  Among these solutions there are 
`one-band' solutions in which  only the function $\psi_{\lambda 
}(k,\nu)$ with  quantum numbers $\lambda = \lambda_0$ and 
$\nu = \nu_0$ is nonzero and all other functions $\psi_{\lambda 
}(k,\nu) = 0$ with $\lambda \neq \lambda_0$ and $\nu \neq \nu_0$. But 
not all of these solutions are stable.

Next we consider the `one-band' self-trapped state which is stable 
and is split off from the lowest energy subband in (\ref{E1234}), 
namely from $E_1(k,0)$ with $\lambda_0 = 1$ and $\nu_0 = 0$.  In this 
case Eq. (\ref{nleqpsi_g}) becomes 
\begin{eqnarray} 
0\, &=&\, \Bigl(E 
- E_{1}(k,0)\Bigr)\psi_{1}(k,0)\nonumber\\ 
&+&\frac{1}{LN}\sum_{k_1,q}
G\left( k,k_1,q \right)
\psi_{1}\sp{\ast}(k_1,0) \psi_{1}(k_1+q,0)\psi_{1}(k-q,0),
\label{nleqpsi_00}
\end{eqnarray}
where
\begin{equation}
G\left( k,k_1,q \right) =
G_{1,1}^{1,1}\left( \begin{array}{ccc}k,&k_1,&q \\ 
0,&0,&0 \end{array} \right) .
\label{G0}
\end{equation}

To solve  (\ref{nleqpsi_00}), we introduce the function
\be
\varphi(\zeta)=\frac{1}{\sqrt {L}} \sum_{k} e^{ikx}
\psi_{1}(k,0)
\label{varphi}
\ee
which depends on the continuous variable $\zeta$, that is a 
dimensionless coordinate along the nanotube axis related to $z$ by 
$\zeta=z/3d$.

Then we assume that in 
the site representation a solution  of (\ref{nleqpsi_00}) is given
 by a wave packet 
broad enough so that it is sufficiently narrow in the $k$- 
representation.  This means that $\psi_{1}(k,0)$ is essentially 
nonzero only in a small region of  $k$-values in the vicinity of 
$k=0$. Therefore, we can use the long-wave approximation
\begin{eqnarray}
E_{1}(k,0)&=&{\cal E}_0-J\sqrt{5 + 4\cos(\frac{k}{2})} \approx
E_1(0) + \frac{1}{12} J k^2 ,\nonumber \\
G\left( k,k_1,q \right) &\approx & G_{0}\left( 0,0,0 \right) = G,
\label{lwappr00}
\end{eqnarray}
where
\be
E_1(0)\, =\,{\cal E}_0\,-\,3J
\label{E01}
\ee 
is the energy bottom of the subband $E_1(k,0)$.
 
Using Eqs. (\ref{F}) - (\ref{W-jj'}) and (\ref{omega}) for
$\nu = \mu =0$ in  the long-wave approximation, 
we can represent the nonlinearity parameter $G$ as
\be
G\, =\,\frac{(\chi_1+G_2)^2 a_1^2+\chi_2^2b_1^2 + 
b_2(\chi_1+G_2)\chi_2 }{k+c^2k_c}
\label{nonlinpar}
\ee
where $a_1$ is a constant of the order of unity, while the constants $b_1, 
b_2, c$ are less than 1. Introducing 
\be 
\Lambda\,=E\,-\,E_1(0),  
\label{Lambda} 
\ee 
we can transform  Eq.(\ref{nleqpsi_00}) into a differential equation for 
$\varphi(\zeta)$:  
\be 
\Lambda \varphi(\zeta) + \frac{J}{12} 
\frac{d^2\varphi(\zeta)}{d \zeta^2} + 
\frac{G}{N}|\varphi(\zeta)|^2 \varphi(\zeta)\, = \,0 ,
\label{nlse} 
\ee 
which is the well-known stationary nonlinear Schr\"{o}dinger equation (NLSE).  
Its normalized solution is given by 
\be 
\varphi(\zeta)=\sqrt{\frac{g_{0} }{2}} 
\frac{1}{\cosh (g_{0} (\zeta-\zeta_0))} 
\label{phi0} 
\ee 
with the eigenvalue 
\be 
\Lambda_{0} = - \frac{J g_0^2 }{12},
\ee
where
\be
g _{0}\,=\,\frac{3 G}{NJ}.
\label{kappa}
\ee
Thus, the eigenenergy of this state is
\be
E_0\, =\, E_1(0) - \frac{3 G^2}{4JN^2} .
\ee

The probability amplitude (\ref{psi_mnj}) of a quasiparticle distribution 
over the nanotube sites, in this state, is given by
\begin{equation}
\psi_{m,n,\varrho} = \frac{1}{2\sqrt{LN}} \sum_{k} e^{ikm}
u_{\varrho,1}(k,0) \psi_{1}(k,0).
\label{fi1}
\end{equation}
The explicit expressions for $u_{\varrho,1}(k,0)$ are given in 
(\ref{etr-coef}).  In the long-wave approximation for the phase 
$\theta _{+}(k,0)$ we find from (\ref{theta}) 
that $\theta _{+}(k,0) \approx k/12$.
Then, using the expressions for $u_{\varrho,1}(k,0)$ and taking into 
account the definition (\ref{varphi}), we obtain 
\begin{equation} 
\psi_{m,n,\varrho} = \frac{1}{2\sqrt{N}} \varphi(z_{m,\varrho}),
\label{fi_j}
\end{equation}
where $z_{m,\varrho}$ are the atom positions along the nanotube axis 
(\ref{Rnmj}):
\begin{eqnarray}
z_{m,1}&=&(m-\frac{1}{3})3d,\qquad z_{m,2}=(m-\frac{1}{6})3d,\nonumber\\
z_{m,3}&=&(m+\frac{1}{6})3d,\qquad z_{m,4}=(m+\frac{1}{3})3d.
\end{eqnarray}

Therefore, according to our solution (\ref{phi0}), the probability
distribution of a quasiparticle over the nanotube sites is given by 
\be
P_\varrho(m,n) = \frac{1}{4N}|\varphi(z_{m,\varrho})|^2=
\frac{g_{0} }{8N} \frac{1}{\cosh ^2(\frac{g_{0}}{3d} z_{m,\varrho})}.
\label{P_0}
\ee
Thus, the quasiparticle is localised along the tube axis and
 distributed uniformly over the tube azimuthal angle.
Therefore, (\ref{P_0}) describes a quasi-1D
large polaron.
In this state, as well as in other one-band states, according 
to (\ref{Q-psi}), only the total symmetrical distortion of the 
nanotube takes place, {\it i.e.}  $Q_{\tau}(q,0) \neq 0$ with $\mu = 
0$ and $Q_{\tau}(q,\mu) = 0$ for $\mu \neq 0$.
The total energy of the polaron state according to (\ref{adiabateq}), is
\be
{\cal E}_0 = W + E_0 = E_1(0) - \frac{G^2}{4JN^2},
\ee
and, thus, depends on the diameter of the nanotube.

\section{Transition to states with broken axial symmetry}

As we see from (\ref{fi_j}),(\ref{phi0}) and (\ref{P_0}), our 
solution, obtained in the long-wave (continuum) approximation, 
possesses the azimuthal symmetry and describes a 
quasi-1D large polaron state which is localized along 
the nanotube axis in the region $\Delta z = \frac{3\pi d}{g_{0}}$. 
Moreover, (\ref{kappa}) shows that as the electron-phonon coupling 
 increases the region of the localization gets smaller.  
Consequently, the wave packet in the quasimomentum representation 
becomes broader and the electron states with higher energies 
participate in the formation of the polaron state. At  strong enough 
coupling the long-wave (continuum) approximation is not valid. 
Moreover, the electron states from the upper bands can also 
contribute to the polaron formation. To consider the transition from 
the large polaron state to the small one, it is convenient to 
transform  Eqs.(\ref{nleqpsi_g}) into the site representation. As a 
first step, let us introduce the functions
\be
\phi_{\varrho}(k,\nu) = \frac{1}{2} \sum_{\lambda} 
  u_{\varrho,\lambda}(k,\nu) \psi_{\lambda }(k,\nu).
\label{phi_jknu}
\ee
Then Eqs.(\ref{nleqpsi_g}) can be rewritten as the following system 
of equations
\begin{eqnarray}
E \phi_{1}(k,\nu) &=&{\cal E}_0 \phi_{1}(k,\nu) - 2J\cos(\frac{\nu}{2})
e^{-i\frac{\nu}{2}} \phi_{2}(k,\nu) - Je^{-ik}\phi_{4}(k,\nu) -
\nonumber\\
&-& \frac{1}{LN} \sum_{k_1,\nu_1,q,\mu} 
\sum_{\varrho_1',\varrho_1,\varrho'} 
T_{1,\varrho'}^{\varrho_1',\varrho_1}(k,\nu;k_1,\nu_1;q,\mu)\phi_{\varrho_1}\sp{\ast}(k_1,\nu_1)
\phi_{\varrho_1'}(k_1+q,\nu_1 +\mu) \phi_{\varrho'}(k-q,\nu -\mu) ,
\nonumber\\
E \phi_{2}(k,\nu) &=&{\cal E}_0 \phi_{2}(k,\nu) - 2J\cos(\frac{\nu}{2})
e^{i\frac{\nu}{2}}\phi_{1}(k,\nu) - J\phi_{3}(k,\nu) -
\nonumber\\
&-& \frac{1}{LN} \sum_{k_1,\nu_1,q,\mu} 
\sum_{\varrho_1',\varrho_1,\varrho'} 
T_{2,\varrho'}^{\varrho_1',\varrho_1}(k,\nu;k_1,\nu_1;q,\mu)\phi_{\varrho_1}\sp{\ast}(k_1,\nu_1)
\phi_{\varrho_1'}(k_1+q,\nu_1 +\mu) \phi_{\varrho'}(k-q,\nu -\mu)  ,
\nonumber\\
E \phi_{3}(k,\nu) &=&{\cal E}_0 \phi_{3}(k,\nu) - 2J\cos(\frac{\nu}{2})
e^{i\frac{\nu}{2}}\phi_{4}(k,\nu) - J\phi_{2}(k,\nu) -
\nonumber\\
&-& \frac{1}{LN} \sum_{k_1,\nu_1,q,\mu} 
\sum_{\varrho_1',\varrho_1,\varrho'} 
T_{3,\varrho'}^{\varrho_1',\varrho_1}(k,\nu;k_1,\nu_1;q,\mu)\phi_{\varrho_1}\sp{\ast}(k_1,\nu_1)
\phi_{\varrho_1'}(k_1+q,\nu_1 +\mu) \phi_{\varrho'}(k-q,\nu -\mu)
\nonumber\\
E \phi_{4}(k,\nu) &=&{\cal E}_0 \phi_{4}(k,\nu) - 2J\cos(\frac{\nu}{2})
e^{-i\frac{\nu}{2}}\phi_{3}(k,\nu) - Je^{ik}\phi_{1}(k,\nu) -
\nonumber\\
&-& \frac{1}{LN} \sum_{k_1,\nu_1,q,\mu} 
\sum_{\varrho_1',\varrho_1,\varrho'} 
T_{4,\varrho'}^{\varrho_1',\varrho_1}(k,\nu;k_1,\nu_1;q,\mu)\phi_{\varrho_1}\sp{\ast}(k_1,\nu_1)
\phi_{\varrho_1'}(k_1+q,\nu_1 +\mu) \phi_{\varrho'}(k-q,\nu -\mu)
\nonumber\\
\label{eq-phi1234}
\end{eqnarray}
where
\begin{equation}
T_{\varrho,\varrho'}^{\varrho_1',\varrho_1}(k,\nu;k_1,\nu_1;q,\mu)  =
\frac{1}{12}\sum_{\tau}
\frac{T_{\varrho,\varrho'}(k-q,\nu-\mu;q,\mu |\tau)T_{\varrho_1',\varrho_1}
\sp{\ast}(k_1,\nu_1;q,\mu |\tau)}
{\omega_{\tau}^{2}(q,\mu)}.
\label{Tfunc}
\end{equation}

In the derivation of these equations we have used the explicit 
expressions (\ref{Gfunc}) and (\ref{F}), the orthonormalization 
conditions (\ref{ortnorm}) and the following expressions for ${\cal 
E}_{\pm}(k,\nu)$
\begin{equation}
{\cal E}_{\pm}(k,\nu) = J\left(2 \cos(\frac{\nu}{2}) \pm e^{-i\frac{k}{2}}  
\right)
e^{\pm 2i\theta_{\pm}(k,\nu)}  =
J\left(2 \cos(\frac{\nu}{2}) \pm e^{i\frac{k}{2}}  \right)
e^{\mp 2i\theta_{\pm}(k,\nu)}.
\end{equation}

To describe the system in the site representation, we introduce
\be
\phi_{\varrho,m}(\nu) = \frac{1}{L} \sum_{k} e^{ikm} \phi_{\varrho}(k,\nu),
\label{phi_jm-nu}
\ee
and obtain
\begin{eqnarray}
E \phi_{1,m}(\nu) &=&{\cal E}_0 \phi_{1,m}(\nu) - 2J\cos(\frac{\nu}{2})
e^{-i\frac{\nu}{2}}\phi_{2,m}(\nu) - J\phi_{4,m-1}(\nu) -
 \frac{a_1^2}{N} \sum_{\nu_1,\mu} \Bigl( 2 \chi_1^2
\phi_{1,m}\sp{\ast}(\nu_1) \phi_{1,m}(\nu_1 +\mu) +
\nonumber\\
&+&  \chi_1 G_2 [
\cos(\frac{\nu_1}{2})e^{i\frac{\nu_1}{2}}
\phi_{2,m}\sp{\ast}(\nu_1) \phi_{1,m}(\nu_1 +\mu) +
\nonumber\\
&+&\cos(\frac{\nu_1 +\mu}{2})e^{-i\frac{\nu_1+\mu}{2}}
\phi_{1,m}\sp{\ast}(\nu_1) \phi_{2,m}(\nu_1 +\mu) +
\phi_{1,m}\sp{\ast}(\nu_1) \phi_{4,m-1}(\nu_1 +\mu) +
\nonumber\\
&+& \phi_{4,m-1}\sp{\ast}(\nu_1) \phi_{1,m}(\nu_1 +\mu)] \Bigr) 
\phi_{1,m}(k-q,\nu -\mu) -
\nonumber\\
&-& \frac{a_1^2}{N} \sum_{\nu_1,\mu}  \Bigl( \chi_1 G_2 [
\phi_{1,m}\sp{\ast}(\nu_1) \phi_{1,m}(\nu_1 +\mu)+ \nonumber\\
&+& 
\cos(\frac{\nu_1 +\mu}{2})e^{-i\frac{\nu_1+\mu}{2}} 
\cos(\frac{\nu_1}{2})e^{i\frac{\nu_1}{2}}
\phi_{2,m}\sp{\ast}(\nu_1) \phi_{2,m}(\nu_1 +\mu)] +
\nonumber\\
&+& 2G_2^2 [
\cos(\frac{\nu_1}{2})e^{i\frac{\nu_1}{2}} \phi_{2,m}\sp{\ast}(\nu_1) 
\phi_{1,m}(\nu_1 +\mu) +
\nonumber\\
&+& \cos(\frac{\nu_1 +\mu}{2})e^{-i\frac{\nu_1+\mu}{2}}
\phi_{1,m}\sp{\ast}(\nu_1) \phi_{2,m}(\nu_1 +\mu) ] \Bigr)
e^{-i\frac{\nu -\mu}{2}} \phi_{2,m}(k-q,\nu -\mu) -
\nonumber\\
&-& \frac{a_1^2}{N} \sum_{\nu_1,\mu}  \Bigl( \chi_1 G_2 [
\phi_{1,m}\sp{\ast}(\nu_1) \phi_{1,m}(\nu_1 +\mu) +
\phi_{4,m-1}\sp{\ast}(\nu_1) \phi_{4,m-1}(\nu_1 +\mu)] +
\nonumber\\
&+& 2G_2^2 [
\phi_{4,m-1}\sp{\ast}(\nu_1) \phi_{1,m}(\nu_1 +\mu) +
\phi_{1,m}\sp{\ast}(\nu_1) \phi_{4,m-1}(\nu_1 +\mu) ]\Bigr)
\phi_{4,m-1}(k-q,\nu -\mu) .
\label{eq-phi_1m}
\end{eqnarray}
with  similar equations for $\varrho=2,3,4$.

When deriving equations (\ref{eq-phi_1m}) we have made a 
qualitative estimate of the expressions of the form
\be
\frac{1}{12}\sum_{\tau}
\frac{W_{\varrho,\varrho'}(k-q,\nu-\mu;q,\mu|\tau) W_{\varrho_1',\varrho_1}^*(k_1,\nu_1;q,\mu|\tau)}
{\omega_{\tau}^{2}(q,\mu)}
\ee
by assuming that the main contribution to these 
quantities comes from the lattice variables with small $q$ and $\mu$.
This gives us an estimate of $a_1$ in Eq.(\ref{eq-phi_1m}).

In zigzag nanotubes, one can identify zigzag chains of carbon atoms which
encircle the nanotube. Let the ring chain  $j$ consists of atoms
enumerated as $(m,n,1)$ and $(m,n,2)$,  the zigzag chain of atoms
$(m,n,3)$ and $(m,n,4)$ corresponds to the ring  $j+1$, and
the chain of atoms $(m-1,n,3)$ and $(m-1,n,4)$, respectively, to the 
ring with number $j-1$. Then we can enumerate atoms as $(j,n,\rho)$ 
where $\rho = 0,1$. Note that the
indices $(j,\rho)$  coincide with the ones used in the numerical 
calculations \cite{us}.

A circle around the nanotube is a zigzag ring chain, with two atoms 
per unit cell, which contains $2N$ atoms. The atoms of the  $j$-th 
chain are equivalent except  that atoms with $\rho = 0$ are coupled 
to the $(j-1)$-th chain and those with $\rho = 1$ to the $(j+1)$-th 
chain, and these two sets of atoms are shifted with respect to each 
other in the opposite directions from some central line, $z_j$, 
(symmetry axes). Thus, we can put 
\begin{eqnarray}
\psi_{j}(\nu ) &=& \frac{1}{\sqrt {2}} \left( \phi_{1,m}(\nu ) + 
e^{-i\frac{\nu}{2}}\phi_{2,m}(\nu) \right)= \frac{1}{\sqrt {2N}} 
\left(\sum _{n=1}^{N-1} e^{-i \frac{\nu}{2}  2n} \phi_{1,m,n} + 
\sum _{n=1}^{N-1} e^{-i \frac{\nu}{2}  (2n+1)} \phi_{2,m,n}
\right)\nonumber\\ 
&=&\frac{1}{\sqrt {2N}} \sum _{l=1}^{2N-1} e^{-i \frac{\nu}{2}  
l} \phi_{m,l}.
\end{eqnarray}

We see that $\psi_{j}(\nu )$ is 
a $k$-representation  for a simple chain 
with $2N$ atoms with $k=\nu/2 =\pi n_1/n $. 
Therefore, this zigzag ring chain is equivalent to 
an isolated nanocircle, studied in \cite{BEPZ_nc}.

Introducing the notation:
$ \phi_{1,m}(\nu ) =\phi_{0,j}(\nu ) ,\ \ \   
e^{-i\frac{\nu}{2}}\phi_{2,m}(\nu) =\phi_{1,j}(\nu ) $ and neglecting 
$\chi_2$ we can rewrite Eq.(\ref{eq-phi_1m}) as follows:  
\be 
E \phi_{0,j}(\nu) \ =\ {\cal E}_0 \phi_{0,j}(\nu) - 
2J\cos(\frac{\nu}{2}) \phi_{1,j}(\nu) - J\phi_{1,j-1}(\nu) - 
\frac{G}{N} \sum_{\nu_1,\mu} 
\phi_{0,j}\sp{\ast}(\nu_1) \phi_{0,j}(\nu_1 +\mu) 
\phi_{0,j}(\nu-\mu),  
\label{e1} 
\ee
where $G$ is given by Eq. (\ref{nonlinpar}). 

For the azimuthal symmetric solution the only nonzero functions are 
those with zero argument,  $\nu =0$. In this case we can use the 
continuum approximation: 
\begin{eqnarray}
\phi_{0,j}&=&\phi (\zeta_{0,j}), \ \ \ \phi_{1,j}=\phi 
(\zeta_{0,j}+\frac{1}{6})=\phi (\zeta_{0,j})+\frac{1}{6} \phi ' 
(\zeta_{0,j}) + \frac{1}{72} \phi '' (\zeta_{0,j}),\nonumber\\ 
\phi_{1,j-1}&=&\phi (\zeta_{0,j}-\frac{1}{3})=\phi 
(\zeta_{0,j})-\frac{1}{3} \phi ' (\zeta_{0,j}) + \frac{1}{18} \phi '' 
(\zeta_{0,j}).
\end{eqnarray}

As a result Eq. (\ref{e1}) transforms into the continuum  NLSE 
(\ref{nlse}). The azimuthally symmetric solution of this equation does not 
always correspond to the state of the lowest energy. To find the 
lowest energy state, we consider Eq. (\ref{e1}) assuming that   
the electron is localized mainly on one chain (for simplicity we label 
it by $ j=0$) and we look for a solution of the form 
\be 
\phi_{\rho ,j}(\nu )=A_{\rho ,j} \phi (\nu ), 
\label{anz} 
\ee 
where 
$A_{\rho ,j}$ are given by Eq. (\ref{phi0}) with $\zeta_0$ describing 
the position of the considered chain.

We can now consider the equation for the 
 chain $j=0$ only. For $\phi (\nu )$ we obtain the 
equation:
\be
(E-{\cal{E}}(\nu )) \phi(\nu )\ +\ 
\frac{G A^2_{0,0} }{N} \sum_{\nu_1,\mu} 
\phi \sp{\ast}(\nu_1) \phi(\nu_1 +\mu) 
\phi (\nu-\mu),  
\label{nls2}
\ee
where
\be
{\cal{E}}(\nu )={\cal{E}}_0-
J\frac{A_{1,-1}}{A_{0,0}}-2J \cos \frac{\nu }{2} . 
\label{en-az}
\ee
Moreover, from (\ref{phi0}) we find that
\be
A_{0,0}=A_{1,0}=\sqrt{\frac{g_0}{2}} \frac{1}{\cosh (g_0/12)},\ \ \ 
A_{1,-1}=\sqrt{\frac{g_0}{2}} \frac{1}{\cosh (5 g_0/12)}. 
\label{az-paa}
\ee

Assuming that the function $\phi (\nu ) $ is essentially nonzero in 
the vicinity of the zero values of $\nu $, the energy dispersion can 
be written in the long-wave approximation as
\be
{\cal E}(k) \ =\ {\cal {E}}(0)\,+\,J\left(\frac{\nu }{2}\right)^2+ 
\dots ,
 \label{en-az1}
\ee
where
\be
{\cal {E}}(0)=\,{\cal {E}}_0\,- \,J\left(2+\frac {A_{1,-1}}{A_{0,0}} 
\right).
\ee

To solve Eq.(\ref{nls2}) we introduce the function
\begin{equation}
\varphi(x)=\frac{1}{\sqrt {2N}} \sum_{k } e^{ikx} \psi(k)
\label{varphi22}
\end{equation}
of the continuum variable $x$ with $k=\nu /2$ being the 
quasimomentum of the nanotube circle with one atom per unit cell. 
Therefore, the quasimomentum representation should be studied in the 
extended band scheme and $-\pi <k< \pi $. Note that $\varphi(x)$ is 
a periodic function, $\varphi(x+2N) = \varphi(x)$, and that the
discrete values of $x=n$, $n = 1,2, ... 2N-1$, 
correspond to the atom positions in the zigzag ring. 

Using the approximation (\ref{en-az1}) one can transform  
(\ref{nls2}) into a nonlinear differential equation for $\varphi 
(x)$ (stationary NLSE):  
\begin{equation} 
J\frac{d^2\varphi(x)}{dx^2}+G A^2_{0,0} |\varphi 
(x)|^2\varphi(x)+\Lambda \varphi(x)=0, 
\label{dnlse} 
\end{equation} 
where $\Lambda = E-{\cal{E}}(0)$.

As it has been shown in \cite{BEPZ_nc}, the solution of Eq. 
(\ref{nls2}), satisfying the normalization condition 
\begin{equation}
\int _0 ^{2N} |\varphi(x)|^2dx=1,
\label{n-cond}
\end{equation}
is expressed via the elliptic Jacobi functions:
\begin{equation}
\varphi(x)=\frac{\sqrt{g}}{2\bf{E}(\it{k})}
dn \left[\frac{2\bf{K}(\it{k})x}{2N},\it{k}  \right].
\label{dn-sol1}
\end{equation}
Here $g=G A^2_{0,0}/(2J) $, and 
$\bf{K}(\it{k})$ and $\bf{E}(\it{k})$ are complete elliptic integrals
of the first and second kind, respectively \cite{BatErd}. The 
modulus of the  elliptic Jacobi function, $\it{k}$,  
is determined from the relation
\begin{equation}
\bf{E}(\it{k})\bf{K}(\it{k}) = \frac{gN}{2}.
\label{leng1}
\end{equation}
The eigenvalue of the solution (\ref{dn-sol1}) is 
\begin{equation}
\Lambda= - J\frac{g^2}{4} \frac{(2-k^2)}{E^2(k)}.
\label{eigen2}
\end{equation}

According to \cite{BEPZ_nc}, the azimuthally symmetric solution 
exists (relation (\ref{leng1}) admits solution) only when $g$ exceeds 
the critical value of the nonlinearity constant
\begin{equation}
g\,>\,g_{cr}\,=\,\frac{\pi^2}{2N},
\label{gcr}
\end{equation}
or, in an explicit form,
\be
\frac{3}{2 \pi ^2}\frac{\sigma ^2}{\cosh ^2(\sigma /(4N))}\ >\ 1, 
\label{gcre} 
\ee
where $\sigma =G/J$ is the adimensional electron-phonon coupling constant.
 
From (\ref{gcre}) we find the critical value of the coupling 
constant, $\sigma _{cr}=2.6$ for $N=8$. According to the numerical 
solution \cite{us}, the critical value of the coupling constant $ 
(\chi _1+G_2)^2/(kJ) \approx 3.2$ for this value of $N$.  Comparing 
this result with the analytical prediction, we conclude, that the 
parameter $a_1$ in (\ref{nonlinpar}) is $a_1 \approx 0.9$.  
Therefore, the estimation  $a_1 \approx 1$ made above is indeed 
valid, which justifies our analytical results.  Of course, the 
applicability of this approach far from the transition breaks down 
because the continuum approximation itself is not valid anymore.

\section{Large polaron states in semiconducting nanotubes}

In zigzag nanotubes, when $N$ is not a multiple of $3$, there is an 
energy gap in the electron spectrum (\ref{E1234}). In the carbon 
SWNT this energy gap, $\Delta$, separates the 1D electron sub-bands 
${\cal E}_0\,-\,{\cal E}_{\pm}(k,\nu)$, which are fully occupied, 
from the empty ones with energy ${\cal E}_0\,+\,{\cal E}_{\pm}(k,\nu)$. 
Such nanotubes are semiconducting \cite{SaiDrDr}. 
Their charge carriers are either electrons (in the conducting band) or holes 
(in the valance band).
For semiconducting zigzag nanotubes $N$ can be 
represented as $N\,=\,3n_0\,+\,1$ or $N\,=\,3n_0\,-\,1$. The lowest 
conducting subband above the energy gap is 
\begin{equation}
E_{3}(k,\nu_0)\,=\,{\cal E}_0\,+\,{\cal E}_{-}(k,\nu _0), 
\end{equation}
and the highest valence subband below the gap is 
\begin{equation}
E_{2}(k,\nu_0)\,=\,{\cal E}_0\,-\,{\cal E}_{-}(k,\nu _0) 
\end{equation}
with $\nu_0 = 2\pi n_0/N$. So, the energy gap in 
semiconducting carbon nanotubes is given by
\be 
\Delta\, = 
E_{3}(0,\nu_0)-E_{2}(0,\nu_0) = 2{\cal E}_{-}(0,\nu _0) = 
\,2J|1-2\cos(\frac{\nu_0}{2})|.
\label{gap}
\ee

Next, we consider a self-trapped state of electrons in the lowest conducting 
band $E_3(k,\nu)$. Because  
${\cal E}_{\pm}(k,-\nu) = {\cal E}_{\pm}(k,\nu),$
this subband is doubly degenerate.
In this case we look for a solution in 
which  only the functions $\psi_{3}(k,\nu)$ with $\nu = \pm \nu_0$ are 
nonzero. Then Eqs. (\ref{nleqpsi_g}) become
\begin{eqnarray}
E\psi _{3}(k,\nu)&=&E_{3}(k,\nu _0)\psi _{3}(k,\nu)- \nonumber \\
&-&\frac{1}{LN}\sum_{k',q,} \Bigl(
\sum_{\nu'} G^{(1)}_{\nu,\nu'}(k,k',q)\,\psi^*_{3}(k',\nu')\psi 
_{3}(k'+q,\nu')\psi _{3}(k-q,\nu) +\nonumber \\ 
&+& G^{(2)}_{\nu,-\nu}(k,k',q)\,\psi^*_{3}(k',-\nu)\psi 
_{3}(k'+q,\nu)\psi _{3}(k-q,-\nu) \Bigr),
\label{nleqpsi3}
\end{eqnarray}
where $\nu,\nu' = \pm \nu _0$ and
\begin{eqnarray}
G^{(1)}_{\nu,\nu'}(k,k',q) &=&G_{3,3}^{3,3}\left( 
\begin{array}{ccc}k,&k',&q \\ \nu,&\nu',&0 \end{array} \right),
\\
G^{(2)}_{\nu,-\nu}(k,k',q) &=& G_{3,3}^{3,3}\left( 
\begin{array}{ccc}k,&k',&q \\ \nu,&-\nu,&2\nu \end{array} \right).
\label{Gs3}
\end{eqnarray}
Here the equivalence of the azimuthal numbers $\mu$ and $\mu \pm 2\pi$ 
should be taken into account.

To solve  (\ref{nleqpsi3}), we introduce functions of the continuum 
variable $x$ using the relation (\ref{varphi})
\be
\varphi_{\nu,3}(x)=\frac{1}{\sqrt {L}} \sum_{k} e^{ikx}
\psi_{3}(k,\nu)
\label{varphi3}
\ee
and use the long-wave approximation
\begin{eqnarray}
E_{3}(k,\nu_0)& \approx & E_3(0,\nu_0)\,+\, \frac{\hbar^2 k^2}{2m}
 ,\nonumber \\
G^{(1)}_{\nu,\nu'}(k,k',q)& \approx & G^{(1)}_{\nu,\nu'}(0,0,0) = G_1,
\nonumber \\
G^{(2)}_{\nu,-\nu}(k,k',q)& \approx & G^{(2)}_{\nu,-\nu}(0,0,0) = G_2.
\label{lwappr33}
\end{eqnarray}
Note that 
\be
E_3(0,\nu_0)\, =\,{\cal E}_0\,+\, \frac{1}{2}\Delta
\label{E03}
\ee
is the energy bottom of the subband $E_3(k,\nu_0)$ and
\be
m\,=\,\frac{2|1-2\cos(\frac{\nu_0}{2})|\hbar^2}{J\cos(\frac{\nu_0}{2})}
\label{m_eff}
\ee
is the quasiparticle effective mass in the subband $E_{3}(k,\nu_0)$.

In this case Eqs.(\ref{nleqpsi3}) are transformed  into a 
differential equations for $\varphi_{\nu,3}(x)$:
\be
\Lambda \varphi _{\nu,3}(x) + \frac{\hbar^2}{2m_{\mu}} \frac{d^2\varphi
_{\mu,3}(x)}{d x^2} + \frac{1}{N} \Bigl(G_1|\varphi _{\nu,3}(x)|^2 +
(G_1+G_2)|\varphi _{-\nu,3}(x)|^2 \Bigr) \varphi _{\nu,3}(x) = 0,
\label{nlse31}
\ee
where
\be
\Lambda\,=E\,-\,E_3(0,\nu_0),
\label{Lambda3}
\ee
and $\nu = \pm \nu_0$.

We observe that equations (\ref{nlse31})
admit two types of soliton-like ground state solutions. The first 
type corresponds to the given azimuthal quantum number: 
$\nu=\nu_0$ or $\nu=-\nu_0$.  Such solutions describe solitons with a
fixed value of the azimuthal number and
 are formed by the electron sublevels  with 
 $\nu$ from the doubly degenerate band, {\it i.e.}, only one 
function $\varphi _{\nu}\neq 0$ is nonzero and the other one  
vanishes:  $\varphi _{-\nu }=0$.  These types of solitons are 
described by the NLSE:  
\be 
\Lambda \varphi _{\nu,3}(x) + 
\frac{\hbar^2}{2m} \frac{d^2\varphi _{\nu,3}(x)}{d x^2} + 
\frac{G_1}{N}|\varphi _{\nu,3}(x)|^2 \varphi _{\nu,3}(x) = 0 .  
\label{nls3} 
\ee 
A normalised solution of this equation is given by 
\be 
\varphi _{3,\nu}(x)=\sqrt{\frac{g_{1} }{2}} \frac{1}{\cosh (g_{1} 
x)} 
\label{f_mu3} 
\ee 
with the eigenvalue 
\be 
\Lambda_{1} = - 
\frac{\hbar^2 g _{1}^2}{2m}, \ee where \be g _{1}=\frac{m 
G_1}{2\hbar^2 N}= \frac{G_1 |1-2\cos(\frac{\nu_0}{2})|}{JN 
\cos(\frac{\nu_0}{2})}.  
\label{g_1} 
\ee 
Thus, the eigenenergy of these states is 
\be 
E_1 = {\cal E}_0\,+\, \frac{1}{2}\Delta \, 
-\,\frac{G_1^2|1-2\cos(\frac{\nu_0}{2})|}{4JN^2\cos(\frac{\nu_0}{2})}.  
\ee

The energy levels of the two solitons with different azimuthal 
numbers ($\nu=\nu_0$ and $\nu=-\nu_0$) are degenerate, similarly to the 
levels of the corresponding bands. However, according to 
Jan-Teller theorem, this degeneracy can be broken by the distortions 
of the lattice resulting in the hybridization of these two states.

Next we consider a case when both these functions are 
nonzero, $\varphi _{\pm} \neq 0 $.  In this case $\varphi _{\pm}$ are 
determined by the system of nonlinear equations (\ref{nlse31}).  A normalised
solution of these equations is given by
\be
\varphi _{\pm\nu_0} = \frac{1}{\sqrt{2}} e^{i\phi_{\pm}}\varphi _{h,3},
\ee
where $\phi_{\pm}$ are arbitrary phases and where $\varphi _{h,3}$ 
satisfies the NLSE (\ref{nls3}) in which the nonlinearity parameter 
$G_1$ is replaced by  $G_1 \longrightarrow (2G_1 + G_2)/2$. Therefore, 
this solution is given by (\ref{f_mu3}) with
\be
g _h= \frac{m (2G_1+G_2)}{4\hbar^2 N} =
 \frac{(2G_1+G_2) |1-2\cos(\frac{\nu_0}{2})|}{2JN \cos(\frac{\nu_0}{2})}.
\label{g_h}
\ee
Its eigenenergy is 
\be
E_h = {\cal E}_0\,+\, \frac{1}{2}\Delta \, 
-\,\frac{(2G_1+G_2)^2|1-2\cos(\frac{\nu_0}{2})|}{16JN^2\cos 
(\frac{\nu_0}{2})}.
\ee

This hybrid soliton possesses a zero azimuthal number while solitons 
(\ref{f_mu3}) have a nonvanishing one: $\nu=\nu_0$ or $\nu=-\nu_0$.  
The energy level of the hybrid soliton state, $E_h$, is lower than 
the level of a soliton with the fixed azimuthal number, $E_1$, 
because $(2G_1+G_2)/2 > G_1$. Note also that the deformation of the 
nanotube in this state is more complicated due to the fact that the 
components $Q_{\pm 2\nu_0}$ of the lattice distortion, as well as the
$Q_0$-component, are non-zero. Moreover, the probability distributions 
of a quasiparticle over the nanotube sites in the state of a hybrid 
polaron and in the state of polaron with a fixed azimuthal number, are 
different.

For a polaron state with a fixed azimuthal quantum number (e.g. 
$\nu=\nu_0$), the probability amplitude (\ref{psi_mnj}) is
\begin{equation}
\psi_{m,n,\varrho} = \frac{1}{2\sqrt{LN}}
\sum_{k}e^{i(km+\nu_0 n)} u_{\varrho,3}(k,\nu_0) \psi_{3}(k,\nu_0).
\label{fi3nu}
\end{equation}
The explicit expressions for $u_{j,3}(k,\nu)$ are given in 
(\ref{etr-coef}).   In the long-wave approximation for the phase $\theta 
_{-}(k,\nu_0)$ we find from (\ref{theta}) that 
\begin{equation}
\tan 2\theta _{-}(k,0)  \approx 
\frac{k}{4(2\cos(\frac{\nu_0}{2})-1)}.
\label{theta_mnu_appr}
\end{equation}
Then, using the expressions for $u_{j,3}(k,\nu)$ and taking into account
the definition (\ref{varphi}), we obtain
\begin{eqnarray}
\psi_{m,n,1}&=&\frac{1}{2\sqrt{N}}
 e^{i\nu_0(n-\frac{1}{4})} \varphi _{\nu,3}(m -\frac{1}{3}+ 
\frac{1}{2}\delta), \nonumber\\ 
\psi_{m,n,2}&=&-\frac{1}{2\sqrt{N}} 
e^{i\nu_0(n+\frac{1}{4})} \varphi _{\nu,3}(m-\frac{1}{6}-\frac{1}{2}\delta),
\nonumber\\
\psi_{m,n,3}&=&-\frac{1}{2\sqrt{N}}
e^{i\nu_0(n+\frac{1}{4})} \varphi _{\nu,3}(m+\frac{1}{6}+ 
\frac{1}{2}\delta),
\nonumber\\
\psi_{m,n,4}&=& \frac{1}{2\sqrt{N}}
e^{i\nu_0(n-\frac{1}{4})} \varphi _{\nu,3}(m+\frac{1}{3}- 
\frac{1}{2}\delta),  
\label{amp3nu} 
\end{eqnarray} 
where 
\be 
\delta = \delta (\nu_0) = 
\frac{\cos(\frac{\nu_0}{2})+1}{3(2\cos(\frac{\nu_0}{2})-1)}.
\label{delta}
\ee

Therefore, according to (\ref{amp3nu}) and (\ref{f_mu3}),
the quasiparticle distribution over the nanotube sites in this 
state is
\be
P_\varrho(m,n) = P(z_{m,\varrho}) = \frac{g_{1} }{8N}
\frac{1}{\cosh ^2(\frac{g_{1}}{3d} (z_{m,\varrho}\pm\frac{1}{2}3d\delta))},
\label{P_3nu}
\ee
where $z_{m,\varrho}$ are the atom positions along the nanotube axis 
(\ref{Rnmj}), the  $``+"$ and $``-"$ signs correspond respectively to atoms 
with an odd ($\varrho=1,3$) and even ($\varrho=2,4$) index $\varrho$ . 
Usually these two types of atoms are labelled as $A$ and $B$ atoms.

We see that the quasiparticle is localised along the tube axis and
is uniformly distributed over the tube azimuthal angle like
a quasi-1D large polaron. But the distributions of the 
quasiparticle among $A$ and $B$ sites are shifted relatively to each 
other by the value $3d\delta (\nu_0)$.

For a hybrid polaron state which possesses zero azimuthal number,
the probability amplitudes (\ref{psi_mnj}) are
\begin{eqnarray}
\psi_{m,n,1}&=&\frac{\cos \left(\nu_0(n-\frac{1}{4}) 
+\phi_0\right)}{\sqrt{2N}} \varphi 
 _{h,3}(m-\frac{1}{3}+\frac{1}{2}\delta),
\nonumber\\
\psi_{m,n,2}&=&-\frac{\cos \left(\nu_0(n+\frac{1}{4}) 
+\phi_0\right)}{\sqrt{2N}} \varphi 
 _{h,3}(m-\frac{1}{6}-\frac{1}{2}\delta),
\nonumber\\
\psi_{m,n,3}&=&-\frac{\cos \left(\nu_0(n+\frac{1}{4}) 
+\phi_0\right)}{\sqrt{2N}} \varphi 
 _{h,3}(m+\frac{1}{6}+\frac{1}{2}\delta),
\nonumber\\
\psi_{m,n,4}&=& \frac{\cos \left(\nu_0(n-\frac{1}{4}) 
+\phi_0\right)}{\sqrt{2N}} \varphi 
 _{h,3}(m+\frac{1}{3}-\frac{1}{2}\delta),
\label{amp3h}
\end{eqnarray}

Therefore, 
the quasiparticle distribution over the nanotube sites in this state is
given by
\be
P_\varrho(m,n) = P(z_{m,\varrho},\phi_{n,\varrho}) = 
  \frac{g_{h} \cos^2 (n_0 \phi_{n,\varrho} +\phi_0) }
       {4N\,\cosh ^2(\frac{g_{h}}{3d}(z_{m,\varrho}\pm\frac{1}{2}3d\delta))},
\label{P_3h}
\ee
where $\phi_{n,\varrho}$ is the angle for the atoms position in the nanotube 
(\ref{Rnmj}), $\phi_{n,\varrho} = n\alpha$ for $\varrho=1,4$ and $\phi_{n,\varrho} 
=(n+\frac{1}{2})\alpha$ for $\varrho=2,3$; $n_0$ is a number which 
determines the azimuthal number $\nu_0$ ($\nu_0 = 2\pi n_0/N$), 
and the $``+"$ and $``-"$ signs correspond to the odd and even values of  
$\varrho$ as above.

We see, that in this polaron state the quasiparticle is localised along 
the tube axis and is modulated over the tube azimuthal angle with the 
angle modulation $2\pi/n_0$.  The longitudinal distributions of the 
quasiparticle among $A$ and $B$ sites are shifted relatively to each 
other by the value $3d\delta (\nu_0)$.

\section{Conclusions}
In this paper we have derived the equations describing self-trapped states
in zigzag nanotubes taking into account the electron-phonon coupling. 
We defined the electron and phonon Hamiltonians using the tight-binding model 
and derived the electron-phonon interaction arising due to the dependence of 
both the on-site and hopping interaction energies on lattice distortions.
Next we performed the adiabatic
approximation and we obtained the basic equations of the model.
These are the equations in the site represantation that were used by us 
in \cite{us} to compute numerical solutions of nanotubes states 
and to determine the ranges of parameters for which
the lowest states were soliton or polaron in nature.

In this paper we have studied this problem analytically. We have shown that
the electrons in low lying states of the electronic Hamiltonian
form polaron-like states.
We have also looked at the sets of parameters for which the continuum 
approximation
holds and the system is described by the nonlinear Schr\"odinger equation.
This has given us good approximate solutions
of the full equations (thus giving us good starting configurations 
for numerical simulations) and has also allowed  
us to compare our predictions with the numerical results \cite{us}.

Our results demonstrate the richness of the spectrum of polaron states.
They include quasi-1D states with azimuthal symmetry for not too 
strong coupling constant, and, at relatively high coupling, states with 
broken azimuthal symmetry which are spread in more than one dimension.
Theoretical estimates of the critical value of the coupling constants between 
the two regimes of self-trapping (with or without axial symmetry) are in good
agreement with our numerical results \cite{us}.

We have also found that for the values of the  parameters corresponding 
to carbon nanotubes, the lowest 
energy states are ring-like in nature 
with their profiles resembling a NLS soliton, {\it ie} similar
to a Davydov soliton as  was claimed in \cite{us}.

We have considered the polaron state of an electron (or a hole)
in semiconducting carbon nanotubes and have shown that the degeneracy
of the conducting (or valence) band with respect to
the azimuthal quantum number plays an important role. The polarons with
lowest energy spontaneously break down the azimuthal symmetry as well
as the translational one and posses an
inner structure: they are self-trapped along the nanotube axis and are
modulated around the nanotube. 

Next we plan to look in more detail at some higher lying states 
and study their properties. We are also planning to study the electric 
conduction properties of our system.

\section{Acknowledgements}
We would like to thank L. Bratek and B. Hartmann for their collaboration 
with us on some topics related to this paper. This work has been supported, 
in part, by a Royal Society travel grant which we gratefully 
acknowledge.

\section{Appendix 1. Diagonalization of the polaron Hamiltonian}

Due to the fact that the diagonal expression
\be
H_{e,0} ={\cal E}_0 \sum_{\ae,\sigma} \,a_{\ae,\sigma} 
\sp{\dagger}a_{\ae,\sigma} 
\ee 
remains diagonal under any unitary transformation, we consider 
only $H_J$. Omitting the multiplier $J$ and the spin index we note
that  $H_J$ is given by, in an explicit form
,
\begin{eqnarray}
H_J &=&-\,\sum_{m,n} 
\Bigl[a_{m,n,1}\sp{\dagger}(a_{m,n-1,2}+a_{m,n,2} +a_{m-1,n,4}) 
\nonumber \\ 
&&+a_{m,n,2}\sp{\dagger} (a_{m,n,1}+a_{m,n,3}+a_{m,n+1,1}) \nonumber 
\\ 
&&+a_{m,n,3}\sp{\dagger} (a_{m,n,4}+a_{m,n+1,4}+a_{m,n,2}) \nonumber 
\\
&&+a_{m,n,4}\sp{\dagger} (a_{m,n-1,3}+a_{m+1,n,1}+a_{m,n,3})  \Bigr].
\label{HJ}
\end{eqnarray}

Due to the translational invariance (with respect to
shifting the index $m$) and 
the rotational invariance (changing $n$) we can perform the 
transformation
\begin{equation}
a_{m,n,\varrho} = \frac{1}{\sqrt{LN}} \sum_{k,\nu}e^{ikm+i\nu n} a_{k,\nu,\varrho},
\label{transf_mn}
\end{equation}
which transforms the Hamiltonian (\ref{HJ}) 
into a diagonal form with respect to the indices $k$ and $\nu$
 and we get
\begin{eqnarray}
H_J &=&-\,\sum_{k,\nu} \Bigl[a_{k,\nu,2}\sp{\dagger} a_{k,\nu,3} +  
a_{k,\nu,3}\sp{\dagger}a_{k,\nu,2}+ \nonumber \\ 
&+& e^{-ik} 
a_{k,\nu,1}\sp{\dagger} a_{k,\nu,4} 
+e^{ik}a_{k,\nu,4}\sp{\dagger}a_{k,\nu,1}+ \nonumber \\ 
&+&2\cos 
\frac{\nu}{2}\left(e^{i\frac{\nu}{2}} a_{k,\nu,3}\sp{\dagger} 
a_{k,\nu,4}+ e^{-i\frac{\nu}{2}} a_{k,\nu,4}\sp{\dagger} a_{k,\nu,3} 
 \right)+ \nonumber \\ 
 &+& 2\cos \frac{\nu}{2}\left( 
e^{-i\frac{\nu}{2}} a_{k,\nu,1}\sp{\dagger}a_{k,\nu,2} + 
 e^{i\frac{\nu}{2}} a_{k,\nu,2}\sp{\dagger} a_{k,\nu,1} \right) \Bigr].
\label{HJ2}
\end{eqnarray}

Note that a direct way to diagonalise (\ref{HJ2}) is to use the 
unitary transformation
\begin{equation}
a_{k,\nu,\varrho} = \frac{1}{2} \sum_{\lambda}u_{\varrho,\lambda}(k,\nu)c_{k,\nu,\lambda}
\label{gtransf}
\end{equation}
with
\begin{equation}
\frac{1}{4} \sum_{\lambda}u_{\varrho,\lambda}\sp{*}(k,\nu)u_{\varrho',\lambda}(k,\nu) = 
   \delta _{\varrho,\varrho'}, \qquad
\frac{1}{4} \sum_{j}u_{\varrho,\lambda}\sp{*}(k,\nu)u_{\varrho,\lambda'}(k,\nu) = 
   \delta _{\lambda,\lambda'}
\label{ortonorm}
\end{equation}
which leads to a system of equations (four in our case) for the coefficients 
$u_{\varrho,\lambda}(k,\nu)$ which 
diagonalise the Hamilionian. A solution of these 
equations would give us the 
coefficients of the transformation as well as the eigenvalues 
$E_{\lambda}(k,\nu)$ 
($\lambda = 1,2,3,4$).  
Instead, we prefer to use a sequential diagonalization.

To do this we choose any two different pairs of operators $a_{k,\nu,\varrho}$ and 
$a_{k,\nu,\varrho'}$ and using unitary transformations we first diagonalise  
two of the four lines in (\ref{HJ2}).  Taking the following 
two pairs:  $\bigl\{a_{k,\nu,1},a_{k,\nu,2} \bigr\}$ and 
$\bigl\{a_{k,\nu,3},a_{k,\nu,4} \bigr\}$, and diagonalising the last 
two lines in (\ref{HJ2}) by the unitary transformations, we get
\begin{eqnarray}
a_{k,\nu,1}&=&\frac{1}{\sqrt{2}}
  \left(e^{-i\frac{\nu}{4}}b_{k,\nu,1}+e^{-i\frac{\nu}{4}}b_{k,\nu,2} \right),
 \nonumber \\
a_{k,\nu,2}&=&\frac{1}{\sqrt{2}}
  \left(e^{i\frac{\nu}{4}}b_{k,\nu,1}-e^{i\frac{\nu}{4}}b_{k,\nu,2} \right),
\label{trI1}
\end{eqnarray}
and
\begin{eqnarray}
a_{k,\nu,3}&=&\frac{1}{\sqrt{2}}
  \left(e^{i\frac{\nu}{4}}b_{k,\nu,3}+e^{i\frac{\nu}{4}}b_{k,\nu,4} \right) ,
 \nonumber \\
a_{k,\nu,4}&=&\frac{1}{\sqrt{2}}
  \left(e^{-i\frac{\nu}{4}}b_{k,\nu,3}-e^{-i\frac{\nu}{4}}b_{k,\nu,4} \right).
\label{trI2}
\end{eqnarray}
Substituting (\ref{trI1}) and (\ref{trI2}) into (\ref{HJ2}), we obtain
\begin{eqnarray}
H_J =&-&\sum_{k,\nu} \bigg\{ 
  \left[ 2\cos \frac{\nu}{2}\left( b_{k,\nu,1}\sp{\dagger}b_{k,\nu,1} +
 b_{k,\nu,3}\sp{\dagger} b_{k,\nu,3} \right) + \cos \frac{k}{2}
\left(e^{-i\frac{k}{2}} b_{k,\nu,1}\sp{\dagger} b_{k,\nu,3}+
 e^{i\frac{k}{2}} b_{k,\nu,3}\sp{\dagger} b_{k,\nu,1} \right)\right]  
  -  \nonumber \\
&-&\left[ 2\cos \frac{\nu}{2}\left(b_{k,\nu,2}\sp{\dagger} b_{k,\nu,2}+
b_{k,\nu,4}\sp{\dagger} b_{k,\nu,4} \right) +
\cos \frac{k}{2}\left( e^{-i\frac{k}{2}} b_{k,\nu,2}\sp{\dagger} b_{k,\nu,4}+
 e^{i\frac{k}{2}} b_{k,\nu,4}\sp{\dagger} b_{k,\nu,2} \right) \right]  
  +   \nonumber \\
&+& i \sin\frac{k}{2} \left( e^{-i\frac{k}{2}} b_{k,\nu,1}\sp{\dagger} 
                            b_{k,\nu,4} -
 e^{i\frac{k}{2}} b_{k,\nu,4}\sp{\dagger}b_{k,\nu,1} 
             - e^{-i\frac{k}{2}} b_{k,\nu,2}\sp{\dagger}b_{k,\nu,3}
 +  e^{i\frac{k}{2}}b_{k,\nu,3}\sp{\dagger}b_{k,\nu,2} \right) \bigg\} .
\label{HJ3}
\end{eqnarray}
Here we have combined the two pairs of operators: 
$\bigl\{b_{k,\nu,1},b_{k,\nu,3} \bigr\}$ with energies $2\cos 
\frac{\nu}{2}$, and $\bigl\{b_{k,\nu,2},b_{k,\nu,4} \bigr\}$ with 
energies $-2\cos \frac{\nu}{2}$. Next we observe that the diagonalization of 
the first two lines in (\ref{HJ3}) reduces to the diagonalization of 
only the non-diagonal parts (the second terms in the square brackets) 
which is achieved  by the transformations similar to 
(\ref{trI1})-(\ref{trI2}):
\begin{eqnarray}
b_{k,\nu,1}&=&\frac{1}{\sqrt{2}}\left(e^{-i\frac{k}{4}}d_{k,\nu,1}
                                     +e^{-i\frac{k}{4}}d_{k,\nu,2} \right),
 \nonumber \\
b_{k,\nu,3}&=&\frac{1}{\sqrt{2}}\left(e^{i\frac{k}{4}}d_{k,\nu,1}
                                     -e^{i\frac{k}{4}}d_{k,\nu,2} \right),
\label{trII1}
\end{eqnarray}
and
\begin{eqnarray}
b_{k,\nu,2}&=&\frac{1}{\sqrt{2}}\left(e^{-i\frac{k}{4}}d_{k,\nu,3}
                                     +e^{-i\frac{k}{4}}d_{k,\nu,4} \right) ,
 \nonumber \\
b_{k,\nu,4}&=&\frac{1}{\sqrt{2}}\left(e^{i\frac{k}{4}}d_{k,\nu,3}
                                     -e^{i\frac{k}{4}}d_{k,\nu,4} \right).
\label{trII2}
\end{eqnarray}
After such transformations, the Hamiltonian (\ref{HJ3}) becomes
\begin{eqnarray}
H_J =&-&\sum_{k,\nu} \bigg\{
\left[ \varepsilon _{+} d_{k,\nu,1}\sp{\dagger}d_{k,\nu,1} -
\varepsilon _{+} d_{k,\nu,3}\sp{\dagger} d_{k,\nu,3} + i \sin \frac{k}{2}
\left(d_{k,\nu,1}\sp{\dagger} d_{k,\nu,3} 
        - d_{k,\nu,3}\sp{\dagger} d_{k,\nu,1} \right)\right] +  \nonumber \\
&+&\left[ \varepsilon _{-} d_{k,\nu,2}\sp{\dagger}d_{k,\nu,2} -
\varepsilon _{-} d_{k,\nu,4}\sp{\dagger} d_{k,\nu,4} - i \sin \frac{k}{2}
\left(d_{k,\nu,2}\sp{\dagger} d_{k,\nu,4} 
        - d_{k,\nu,4}\sp{\dagger} d_{k,\nu,2} \right)\right]
\label{HJ4}
\end{eqnarray}
where
\begin{equation}
\varepsilon _{+}=\varepsilon _{+}(k,\nu) = 2\cos \frac{\nu}{2} 
              +  \cos \frac{k}{2}, \quad
\varepsilon _{-}=\varepsilon _{-}(k,\nu)= 2\cos \frac{\nu}{2} 
              - \cos \frac{k}{2} .
\label{eps_pm}
\end{equation}

Thus we have obtained two independent pairs of operators: 
$\bigl\{d_{k,\nu,1},d_{k,\nu,3} \bigr\}$ with energies $\varepsilon 
_{+}$ and $-\varepsilon _{+}$, and $\bigl\{d_{k,\nu,2},d_{k,\nu,4} 
\bigr\}$, with energies  $\varepsilon _{-}$ and $-\varepsilon _{-}$.  
So, the diagonalization of the Hamiltonian (\ref{HJ}) is reduced to 
the diagonalization of two independent quadratic forms. The first and second  
lines in (\ref{HJ4}) are diagonalised respectively by the unitary 
transformation
\begin{eqnarray}
d_{k,\nu,1}&=&\cos \theta _{+}c_{k,\nu,1} - i\sin \theta _{+} c_{k,\nu,4} ,
 \nonumber \\
d_{k,\nu,3}&=&-i \sin \theta _{+}c_{k,\nu,1} + \cos \theta _{+} c_{k,\nu,4}
\label{trIII1}
\end{eqnarray}
and 
\begin{eqnarray}
d_{k,\nu,2}&=&\cos \theta _{-}c_{k,\nu,2} + i\sin \theta _{-} c_{k,\nu,3} ,
 \nonumber \\
d_{k,\nu,4}&=&i \sin \theta _{-}c_{k,\nu,2} + \cos \theta _{-} c_{k,\nu,3}.
\end{eqnarray}
Here $\theta _{\pm} = \theta _{\pm}(k,\nu)$ are determined from the 
relations
\begin{equation}
 \tan 2\theta _{\pm} = \frac{\sin \frac{k}{2}}{2\cos \frac{\nu}{2} 
                     \pm \cos \frac{k}{2}} .
\label{theta}
\end{equation}

After this we obtain the final 
expression for the Hamiltonian (\ref{HJ}) in the diagonal 
representation:
\begin{eqnarray}
H_J = \sum_{k,\nu} \left[ -{\cal E}_{+} c_{k,\nu,1}\sp{\dagger}c_{k,\nu,1} -
{\cal E}_{-} c_{k,\nu,2}\sp{\dagger} c_{k,\nu,2} +
{\cal E}_{-} c_{k,\nu,3}\sp{\dagger} c_{k,\nu,3} +
{\cal E}_{+} c_{k,\nu,4}\sp{\dagger} c_{k,\nu,4} \right],
\label{HJ5}
\end{eqnarray}
where
\begin{eqnarray}
{\cal E}_{\pm}&=& {\cal E}_{\pm} (k,\nu)=\varepsilon _{\pm}\cos 2\theta _{\pm}
+ \sin \frac{k}{2} \sin 2\theta _{\pm} = \nonumber \\
&=& \sqrt{\varepsilon _{\pm}^2 + \sin^2 \frac{k}{2}}=
\sqrt{1+4\cos^2 \frac{\nu}{2} \pm 4\cos \frac{\nu}{2} \cos \frac{k}{2}} .
\label{E_pm}
\end{eqnarray}
Thus, the electron Hamiltonian (\ref{Hamilt1_mn}) has been transformed 
into the diagonalised form (\ref{Hamilt2})
\begin{equation}
H_e= H_{e,0}+H_J =\,\sum_{k,\nu,\lambda,\sigma} E_{\lambda}(k,\nu)\,
c_{k,\nu,\lambda,\sigma}\sp{\dagger}c_{k,\nu,\lambda,\sigma}
\label{Hel6}
\end{equation}
with the energy bands (\ref{E1234}).

Combining all the transformations together, we can write down the resulting 
unitary transformation:
\begin{eqnarray}
a_{k,\nu,1}&=&\frac{1}{2} e^{-i\frac{k+\nu}{4}}\left(e^{-i\theta_+}c_{k,\nu,1}
+e^{i\theta_-}c_{k,\nu,2}+ e^{i\theta_-}c_{k,\nu,3})
                         + e^{-i\theta_+}c_{k,\nu,4}\right),
\nonumber\\
a_{k,\nu,2}&=&\frac{1}{2} e^{-i\frac{k-\nu}{4}}\left(e^{i\theta_+}c_{k,\nu,1}
+e^{-i\theta_-}c_{k,\nu,2}- e^{-i\theta_-}c_{k,\nu,3})
                         - e^{i\theta_+}c_{k,\nu,4}\right),
\nonumber\\
a_{k,\nu,3}&=&\frac{1}{2} e^{i\frac{k+\nu}{4}}\left(e^{-i\theta_+}c_{k,\nu,1}
-e^{i\theta_-}c_{k,\nu,2}- e^{i\theta_-}c_{k,\nu,3})
                         + e^{-i\theta_+}c_{k,\nu,4}\right),
\nonumber\\
a_{k,\nu,4}&=&\frac{1}{2} e^{i\frac{k-\nu}{4}}\left(e^{i\theta_+}c_{k,\nu,1}
-e^{-i\theta_-}c_{k,\nu,2}+ e^{-i\theta_-}c_{k,\nu,3})
                         - e^{i\theta_+}c_{k,\nu,4}\right)
\label{tr-fin}
\end{eqnarray}
which can be written in the general form (\ref{gtransf}).

The inverse transformation is
\begin{eqnarray}
c_{k,\nu,1}&=&\frac{1}{2}\bigg( 
        e^{i\left(\frac{k+\nu}{4}+\theta_+\right)}a_{k,\nu,1}
+e^{i\left(\frac{k-\nu}{4}-\theta_+\right)}a_{k,\nu,2}
+ e^{-i\left(\frac{k+\nu}{4}-\theta_+\right)}a_{k,\nu,3}
+e^{-i\left(\frac{k-\nu}{4}+\theta_+\right)}a_{k,\nu,4}\bigg),
\nonumber\\
c_{k,\nu,2}&=&\frac{1}{2}\bigg( 
        e^{i\left(\frac{k+\nu}{4}-\theta_-\right)}a_{k,\nu,1}
+e^{i\left(\frac{k-\nu}{4}+\theta_-\right)}a_{k,\nu,2}
- e^{-i\left(\frac{k+\nu}{4}+\theta_-\right)}a_{k,\nu,3}
-e^{-i\left(\frac{k-\nu}{4}-\theta_-\right)}a_{k,\nu,4}\bigg),
\nonumber\\
c_{k,\nu,3}&=&\frac{1}{2}\bigg( 
        e^{i\left(\frac{k+\nu}{4}-\theta_-\right)}a_{k,\nu,1}
-e^{i\left(\frac{k-\nu}{4}+\theta_-\right)}a_{k,\nu,2}
- e^{-i\left(\frac{k+\nu}{4}+\theta_-\right)}a_{k,\nu,3}
+e^{-i\left(\frac{k-\nu}{4}-\theta_-\right)}a_{k,\nu,4}\bigg),
\nonumber\\
c_{k,\nu,4}&=&\frac{1}{2}\bigg( 
        e^{i\left(\frac{k+\nu}{4}+\theta_+\right)}a_{k,\nu,1}
-e^{i\left(\frac{k-\nu}{4}-\theta_+\right)}a_{k,\nu,2}
+ e^{-i\left(\frac{k+\nu}{4}-\theta_+\right)}a_{k,\nu,3}
-e^{-i\left(\frac{k-\nu}{4}+\theta_+\right)}a_{k,\nu,4}\bigg)\nonumber\\
\end{eqnarray}
which can be written in the general form
 as
\begin{equation}
c_{k,\nu,\lambda} = \frac{1}{2} \sum_{\varrho}u_{\varrho,\lambda}\sp{*}(k,\nu)a_{k,\nu,j}.
\label{invgtr}
\end{equation}

\section{Appendix 2. Semi-Classical Equations}

Due to the fact that $H_{na}$ (\ref{Hna}) is nondiagonal, its average value on 
the 
wavefunctions (\ref{adappr}) vanishes: $\langle\Psi |H_{na}|\Psi\rangle
= 0 $. The average value of the total Hamiltonian, 
${\cal H} = \langle\Psi |H |\Psi\rangle =E,$  
gives us the energy in the zero adiabatic approximation.  The calculation of
${\cal H}$ with (\ref{uoperat}) and
(\ref{adappr}) gives us the Hamiltonian functional of classical displacements
$\vec{U}_{\ae}$ and the quasiparticle wavefunction $ 
\varphi_{i,j,\rho}$.  So we see that the zero adiabatic approximation 
leads to the semiclassical approach which is often used in the 
description of self-trapped states. 

Calculating ${\cal H}$, we get the Hamiltonian functional, 
\begin{eqnarray}
{\cal H} &=&\,H_{ph}\, +\,\sum_{\ae} \Bigl({\cal E_0}\,\cmod{\ph}\,
- J\,\sum_{\delta }\varphi _\ae^*\varphi_{\delta (\ae)}
+\chi_1\,\cmod{\ph}\, \sum_{\delta } W\delta_\ae 
\nonumber\\
&+&\chi_2\,\cmod{\ph}\,C_{\ae}
+ G_2\sum_\delta  
\ph^*\varphi_{\delta (\ae)}W\delta _\ae
\Bigr).
\label{EP-hamiltonian}
\end{eqnarray}
Here $H_{ph}$ is given by (\ref{phon-ham1}), where the displacements 
and the canonically conjugate momenta are classical variables. The 
site labelling in (\ref{EP-hamiltonian}) corresponds to the 
elementary cell with two atoms, based on the non-orthogonal basis 
vectors, and index $\ae$ labeles sites $i,j,\rho $.

From (\ref{EP-hamiltonian}) we derive the following static equations 
for the functions $u$, $v$, $s$ and $\ph$: 
\begin{eqnarray}
0 &=&({\cal W }+{\cal E_0})\pc - J\,\bigl(\pr+\pl+\pd\bigr)
+\chi_1\,\pc\,(W_r+W_l+W_d)+ \chi_2\,\pc\,C\nonumber\\
&+&G_2[\pr W_r +\pl W_l+\pd W_d],
\label{EqnPhi2}
\end{eqnarray}
\begin{eqnarray}
0 &=&k \Bigl[ \sqrt{3} \cos(\frac{\alpha}{4})(W_l-W_r)
             + \cos(\frac{\alpha}{4})(\Omega_l-\Omega_r) + 2 \Omega_d
\Bigr] 
  + k_c \Bigl[ \sin(\frac{\alpha}{4})(\frac{5}{2} \cos^2(\frac{\alpha}{4})-1)
       (C_l -C_r) 
\Bigr] \nonumber\\
&+&\chi_1 (\cmod{\pl}-\cmod{\pr})
        (\frac{\sqrt{3}}{2} \cos(\frac{\alpha}{4}))
 +\chi_2 (\cmod{\pl}-\cmod{\pr})
           \sin(\frac{\alpha}{4})(\frac{5}{2} \cos^2(\frac{\alpha}{4})-1)
\nonumber\\
&+& {\sqrt{3}\over 2} \cos(\frac{\alpha}{4}) G_2 
   (\pc^*\pl+\pl^*\pc-\pc^*\pr+\pr^*\pc),
\label{Eqnu2}
\end{eqnarray}
\begin{eqnarray}
0 &=&k \Bigl[ (2 W_d -W_r-W_l) +\sqrt{3}(\Omega_r+\Omega_l)
\Bigr] 
  + k_c \Bigl[ \frac{\sqrt{3}}{4}\sin(\frac{\alpha}{2})
        ( 2 C + C_r + C_l)
\Bigr] \nonumber\\
&+&\chi_1 \frac{1}{2} (2\cmod{\pd}-\cmod{\pr}-\cmod{\pl})
+\chi_2 (2\cmod{\pc}+\cmod{\pr}+\cmod{\pl})
        (\frac{\sqrt{3}}{4}\sin(\frac{\alpha}{2}) )
\nonumber\\
&+& \frac{1}{2} G_2 
   (-\pc^*\pr-\pr^*\pc-\pc^*\pl-\pl^*\pc + 2\pc^*\pd+2\pd^*\pc ),
\label{Eqnv2}
\end{eqnarray}
\begin{eqnarray}
0 &=&k \Bigl[ \sqrt{3} \sin(\frac{\alpha}{4}) (Wr+Wl )
             +\sin(\frac{\alpha}{4}) (\Omega_r +\Omega_l )
       \Bigr] \nonumber\\
  &+& k_c \Bigl[
(\frac{3}{2}\cos(\frac{\alpha}{4})-\frac{5}{2}\cos^3(\frac{\alpha}{4}))
   (C_r+C_l) 
 -\cos(\frac{\alpha}{4})C_d + 3\cos^3(\frac{\alpha}{4})C
\Bigr] \nonumber\\
&+&\chi_1 \frac{\sqrt{3}}{2} \sin(\frac{\alpha}{4}) 
     (2\cmod{\pc}+\cmod{\pr}+\cmod{\pl})\nonumber\\
&+&\chi_2 \Big(
(\frac{3}{2}\cos(\frac{\alpha}{4})-\frac{5}{2}\cos^3(\frac{\alpha}{4}))
   (\cmod{\pr}+\cmod{\pl}) 
-\cos(\frac{\alpha}{4})\cmod{\pd}
   + 3\cos^3(\frac{\alpha}{4})\cmod{\pc}
   \Bigr)\nonumber\\
&+& G_2 {\sqrt{3}\over 2} \sin(\frac{\alpha}{4})
     (\pc^*\pr+\pr^*\pc+\pc^*\pl+\pl^*\pc).
\label{Eqns2}
\end{eqnarray}

These equations were used in \cite{us}
to determine numerically the conditions for the existence
of polaron/soliton states.

{}


\begin{thebibliography}{99}

\bibitem{SaiDrDr}
{R. Saito, G. Dresselhaus and M.S. Dresselhaus, 
{\it Physical Properties of Carbon Nanotubes} 
(Imperial College Press, London, 1998).}
\bibitem{DrDrEkl}
{M.S. Dresselhaus, G. Dresselhaus and P.C. Ekland, 
{\it Science of Fulerens and Carbon Nanotubes} (Academic, New York, 1996).}
\bibitem{Dai}
H. Dai, Surf. Sci. {\bf 500}, 208 (2002).
\bibitem{Benny}
{T. Hemraj-Benny, S. Bannerjee, et al., Phys. Chem. Chem. Phys. {\bf 7}, 
1103 (2005).}
\bibitem{nanosw}
{J. E. Jang, S. N. Cha, Y. Choi,  et al., 
Appl. Phys. Lett. {\bf 87}, 163114 (2005).}
%J. E. Jang, S. N. Cha, Y. Choi, G. A. J. Amaratunga, D. J. Kang, 
%D. G. Hasko, J. E. Jung and  J.  M.  Kim ,
%Nanoelectromechanical switches with vertically aligned carbon 
%nanotubes
\bibitem{Duc}
L. Duclaux, Carbon {\bf 40}, 1751 (2002).
\bibitem{WoMah}
L.M. Woods and G.D. Mahan, Phys. Rev. B {\bf 61},10 651 (2000).
\bibitem{Mah}
G.D. Mahan, Phys. Rev. B {\bf 68}, 125409 (2003).
%Electron-optical phonon interaction in carbon nanotubes
\bibitem{SFDrDr}
R. Saito, M. Fujita, G. Dresselhaus and M.S. Dresselhaus, Appl. Phys. Lett. 
{\bf 60} 2204 (1992).
\bibitem{MDWh}
J.W. Mintmire, B.I. Dunlap, and C.T. White, Phys. Rev. Lett. {\bf 68}, 
631 (1992).
\bibitem{Kane}
{C.L. Kane, E.J. Mele, Phys. Rev. Lett., {\bf 78}, 1932 (1997).}
%Size, shape and low-energy electronic structure of carbon nanotubes. 
%pp. 1932-1935.
\bibitem{Chamon}
{C. Chamon, Phys. Rev. B {\bf 62}, 2806 (2000).}
%Solitons in carbon nanotubes, pp. 2806-2812.
\bibitem{Alves}
M. Verissimo-Alves, R.B. Capaz, B. Koiller, E. Artacho, and H. Chacham,
Phys. Rev. Lett. {\bf 86}, 3372 (2001).
%Polarons in carbon nanotubes
\bibitem{JiDrDr}
R.A. Jishi, M.S. Dresselhaus, G. Dresselhaus, Phys. Rev. B {\bf 48}, 
11385 (1993).
\bibitem{PStZ}
L. Pietronero, S. Str{\"a}ssler, H.R. Zeller, M.J. Rice, Phys. Rev. B 
{\bf 22}, 904 (1980).
\bibitem{Dav}
A.S. Davydov, {\it Solitons in Molecular Systems} (Dordrecht,
Reidel, 1985).
\bibitem{Peierls}
{R.E. Peierls, {\it Quantum Theory of Solids} (Clarendon, Ozford, 1955).}
\bibitem{Froehlich}
{H. Froehlich, Proc. Royal Soc. London Ser. A {\bf 215}, 291 (1952).}
\bibitem{Furer}
M.S. Fuhrer, B.M. Kim, T. Durkop, T. Brintlinger, Nano Lett. {\bf 2}, 
755 (2002);
%High-mobility nanotube transistor memory
T. Durkop, S.A. Getty, E. Cobas, and M.S. Fuhrer, {\it ibid} {\bf 4}, 
35 (2004).
%Extraordinary mobility in semiconducting carbon nanotubes
\bibitem{Rados}
M. Radosavljevic, M. Freitag, K.V. Thadami, and A.T. Johnson, Nano Lett. 
{\bf 2}, 761 (2002).
%Nonvolatile molecular memory elements based on ambipolar nanotube field 
%effect transistors
\bibitem{Pryl}
Yu. Prylutskyy, A. Suprun, O. Ogloblya, in {\it Book of abstracts} of 
Intern. Conference on Theoretical Physics 
(Paris, UNESCO, July 22nd-27th, 2002) P.278;
%Solitons in carbon nanotube
Yu.I. Prylutskyy, O.V. Ogloblya, P. Eklund, and P. Scharff, Synth. Met. (2001).
%Electronic properties of carbon nanotubes with defects.
\bibitem{SSH}
W.P. Su, J.R. Schriefer, A.J. Heeger, Phys. Rev. Lett. {\bf 42}, 1698 (1070);
Phys. Rev. B {\bf 22}, 2099 (1980).
\bibitem{us} 
L. Bratek, L. Brizhik, A. Eremko, B. Piette, M. Watson
and W. Zakrzewski, preprint cond-mat/0510338
\bibitem{Kempa}
K. Kempa, Phys. Rev. B {\bf 66}, 195406 (2002).
\bibitem{BEPZ_nc}
L.S. Brizhik, A.A. Eremko, B..Piette and W. Zakrzewski  
%Electron self-trapping on a nano-circle
Arxiv preprint cond-mat/0503366, (2005)
\bibitem{Maradudin}
 {A.A. Maradudin, E.W. Montroll, G.H. Weiss,
 {\it Theory of Lattice Dynamics in the Harmonic Approximation}
 (Academic Press, New York and London, 1963).} 
\bibitem{Sanch}
D. Sanchez-Portal, E. Artacho, J.M. Soler, A. Rubio, and P. Orddejon,
Phys. Rev. B {\bf 59}, 12678 (1999).
\bibitem{Mah_vibr}
G.D. Mahan, Phys. Rev. B {\bf 65}, 235402 (2002).
\bibitem{BatErd}
H. Bateman and A. Erd{\'e}lyi, {\it Higher Transcendental Functions}
(McGraw-Hill, New York, 1955), Vol.3.


\bibliographystyle{plain}
\end{thebibliography}
\end{document}